\documentclass[aps,prc,superscriptaddress,notitlepage,nofootinbib]{revtex4-1}

\usepackage{epsfig}
\usepackage{color}
\usepackage[latin1]{inputenc}
\usepackage{float,amsmath}
\usepackage{graphicx}

\newcommand{\be}{\begin{eqnarray}}
\newcommand{\ee}{\end{eqnarray}}

\newcommand{\benum}{\begin{enumerate}}
\newcommand{\eenum}{\end{enumerate}}

\begin{document}

\title{Viscous photons in relativistic heavy ion collisions}

\author{Maxime Dion}
\affiliation{Department of Physics, McGill University, 3600 University
Street, Montreal, Quebec, H3A\,2T8, Canada}
\author{Jean-Fran\c{c}ois Paquet}
\affiliation{Department of Physics, McGill University, 3600 University
Street, Montreal, Quebec, H3A\,2T8, Canada}
\author{Bj\"orn Schenke}
\affiliation{Physics Department, Bldg. 510A, Brookhaven National
Laboratory, Upton, NY 11973, USA}
\author{Clint Young}
\affiliation{Department of Physics, McGill University, 3600 University
Street, Montreal, Quebec, H3A\,2T8, Canada}
\author{Sangyong Jeon}
\affiliation{Department of Physics, McGill University, 3600 University
Street, Montreal, Quebec, H3A\,2T8, Canada}
\author{Charles Gale}
\affiliation{Department of Physics, McGill University, 3600 University
Street, Montreal, Quebec, H3A\,2T8, Canada}

\begin{abstract}
Theoretical studies of the production of real thermal photons  in relativistic heavy
ion collisions at the Re\-lativistic Heavy Ion Collider (RHIC)  are performed.  The space-time
evolution of the colliding system is modelled using  \textsc{music}, a 3+1D
relativistic hydrodynamic simulation{, using both its ideal and viscous versions}. The inclusive spectrum and its
azimuthal angular anisotropy are studied separately, and the relative contributions 
of the different photon sources are highlighted. It is shown that the
photon $v_2$ coefficient is especially sensitive to the details of the microscopic dynamics like the  equation of state, the ratio of shear
viscosity over entropy density, $\eta/s$, and to the morphology of the initial
state.
\end{abstract}

\maketitle

\section{Introduction}
The study of relativistic collisions of nuclei constitutes a vibrant branch
of subatomic physics that straddles nuclear and particle physics. It offers
a privileged window on the physics of hot and dense strongly interacting
matter and as such, it complements astrophysical studies. There, the
hadronic equation of state is an ingredient of paramount importance that
enters the evaluation of the bulk properties of neutron stars, for
example. In comparison, relativistic nuclear collisions do offer the
considerable practical advantage of providing laboratory control over the
projectile and target characteristics, together with the beam energy. This
physics currently defines a large experimental effort being pursued at 
the Relativistic Heavy Ion Collider (RHIC), at Brookhaven National Laboratory 
and, more recently, at the Large Hadron Collider (LHC), at CERN. One of
the remarkable results that emerged from the RHIC program so far, is the
fact that the hot and dense hadronic matter produced there
\cite{Adcox:2004mh,Adams:2005dq,Back:2004je,Arsene:2004fa} could be
described using almost ideal  hydrodynamics, that is with a small shear
viscosity coefficient $\eta$ \cite{Kolb:2003dz},  compared to the entropy density $s$. The first LHC flow results
for heavy ion collisions \cite{Aamodt:2010pa,*Steinberg:2011dj} also
suggest similar conclusions: a recent overview of flow results can be found
in Ref. \cite{[{See, for example, }][{, and references
therein.}]Schenke:2011qd}. In fact, the progress in both theoretical and in
experimental analyses has been such that the goal of a quantitative
extraction of the shear viscosity coefficient of hot and dense strongly
interacting matter from relativistic nuclear collision data now appears
closer than ever
\cite{Schenke:2010rr,Schenke:2011tv,Shen:2011zc,Song:2011qa,Luzum:2008cw,Romatschke:2007mq}.

In heavy-ion collisions, the flow has been characterized by considering a
Fourier expansion of the the triple differential cross-section, with the
variable being the azi\-muthal angle with respect to the reaction plane
\cite{Poskanzer:1998yz}:
\begin{eqnarray}
E \frac{d^3 N}{d^3 p} = \frac{1}{2 \pi} \frac{d^2N}{p_T d p_T d y}\left(1 +
\sum_{n = 1}^\infty 2 v_n \cos \left[n \left( \phi - \psi_r \right) \right]
\right)\nonumber \\
\label{v_n}
\end{eqnarray}
where $\psi_r$ is the reaction plane angle. The expansion coefficients,
$v_n$, will then quantify the degree of azimuthal anisotropy. In the
progress towards a precise and quantitative characterization of the
hydrodynamical state of nuclear collisions, a recent development consisted of linking the odd-numbered coefficients to fluctuations in the
initial state \cite{Alver:2010gr}. Indeed, up to this point the flow
observables in nuclear collisions had been analyzed by considering the
coefficients of the Fourier expansion of the azimuthal angle distribution
of the particle spectra, together with smooth initial conditions for which
the odd values of the Fourier coefficients vanish \cite{Poskanzer:1998yz}.
It is fair to write that a rich and quantitative picture of nuclear flow is now emerging\footnote{{ A discussion on how to calculate the elliptic flow, $v_2$, in the presence of fluctuating initial conditions appears later in this paper.}}.

 In general, the measured hadronic observables give a dynamical snapshot of
the conditions that existed on the freeze-out hypersurface. In contrast,
electromagnetic radiation is emitted throughout the space-time evolution
and suffers negligible final-state interactions, owing mainly to the
smallness of $\alpha$, the electromagnetic coupling constant. Real and
virtual photons are thus penetrating probes, and as such can carry
information about the different stages of the high energy collisions. A
consequence of this statement is that accurate and meaningful calculations
of  photon spectra in relativistic nuclear collisions will need realistic
electromagnetic emissivities and precise modelling of the space-time
dynamics. The goal in this paper is to extend the calculations of real
photon production to situations which incorporate the developments made on
the purely hadronic front. Cases where the emitting source is no longer in
local thermal equilibrium will be considered, together with cases where the
initial states of the nucleus-nucleus collisions are no longer smooth but
are allowed to fluctuate event-by-event. The paper is organized as follows:
section \ref{hydro_evol} contains a brief description of \textsc{music}, 
our implementation of 3+1 viscous hydrodynamics which is used to calculate the evolution of the background medium.
In section \ref{viscous}, we give a short explanation of viscous corrections to the
local momentum distribution function and the photon emission rates used in
this study. Our main results are presented in section \ref{results} and we conclude in
section \ref{conclusion}.

\section{Hydrodynamical evolution}
\label{hydro_evol}

As mentioned earlier, photons are penetrating probes that are emitted
throughout the heavy-ion collision. It is thus imperative to  evaluate
their observed properties with a time-evolution scenario that is both realistic and
consistent with a large number of empirical observables. One such approach
is \textsc{music}, a three-dimensional simulation of relativistic
hydrodynamic systems \cite{Schenke:2010nt}. The general features of
\textsc{music} are described below, first in the ideal limit, and then
incorporating a finite coefficient of shear viscosity.

The solution of the conservation laws for the stress-energy tensor and the
net baryon current, $T^{\mu \nu}$ and $J_{\rm B}^\mu$, respectively,
dictate the evolution in time of an ideal hydrodynamical system.
More specifically,
\begin{eqnarray}
\partial_\mu T^{\mu \nu}_{\rm ideal} &= &0\,, \nonumber \\
\partial_\mu J_{\rm B, ideal}^\mu &= &0
\label{conser}
\end{eqnarray}
and
\begin{eqnarray}
T_{\rm ideal}^{\mu \nu} &=& \left( \epsilon + P\right) u^\mu u^\nu - P
g^{\mu \nu}, \nonumber \\
J_{\rm B, ideal} ^\mu &=& \rho_{\rm B} u^\mu
\end{eqnarray}
Note that $P$ is the local pressure, $\epsilon$ is the local energy
density, $\rho_{\rm B}$ is the local net baryon density,  $u^\mu =
\left(\gamma, \gamma {\bf v}\right)$ is the local flow velocity with
respect to some fixed frame, and $g^{\mu \nu} = {\rm diag}\left(1, -1, -1,
-1\right)$. This represents a set of five scalar equations, with six
unknowns. The set is closed by specifying an equation of state,
$P(\epsilon, \rho_{\rm B})$. \textsc{music} is implemented in $\tau -
\eta_s$ coordinates, where $\tau$ is the proper time, and $\eta_s$ the
space-time rapidity. The transformations to real time  and longitudinal
coordinate variables, $\{ t, z\}$, are 
\begin{eqnarray}
t = \tau \cosh \eta_s, \ \ z = \tau \sinh \eta_s
\end{eqnarray}
The solution of Eqs. (\ref{conser}) in $\tau-\eta_s$ coordinates is
obtained with the Kurganov-Tadmor method \cite{Kurganov:2000}, and an
equation of state extracted from lattice QCD calculations
\cite{Huovinen:2009yb} is used in this work. Importantly, \textsc{music} is
a three-dimensional simulation, and is therefore capable of following the
time evolution of the rapidity profile. It has been used for the successful
calculation of flow variables, including elliptic flow and higher flow
harmonics \cite{Schenke:2010nt}. A discussion of results with ideal
hydrodynamics is postponed, turning now to the inclusion of viscous
effects.

The first-order - or Navier-Stokes - formalism for viscous hydrodynamics is
known to introduce unphysical superluminal signals that spoil the theory's
stability. Various formulations of second-order hydrodynamics
\cite{Israel:1976tn,Stewart:1977,Israel:1979wp,Grmela:1997zz,Muronga:2001zk} address this problem, and a variant \cite{Baier:2007ix} of the Israel-Stewart formalism is used here. In this approach, the stress-energy tensor is $T^{\mu \nu} = T_{\rm ideal}^{\mu \nu} + \pi^{\mu \nu}$, and the evolution equation are
\begin{eqnarray}
\partial_\mu T^{\mu \nu} & = & 0\,, \nonumber \\
\Delta^\mu_\alpha \Delta^\nu_\beta u^\sigma \partial_\sigma \pi^{\alpha
\beta}& =& - \frac{1}{\tau_\pi} \left(\pi^{\mu \nu} - S^{\mu \nu} \right) -
\frac{4}{3} \pi^{\mu \nu} \left( \partial_\alpha u^\alpha \right) \nonumber
\\
\label{W_evol}
\end{eqnarray}
where $\Delta^{\mu \nu} = g^{\mu \nu} - u^\mu u^\nu$, and $\tau_\pi$ is
usually interpreted as a relaxation time . The first-order (in velocity gradients) viscous part of the
stress-energy tensor appears here and is
\begin{eqnarray}
S^{\mu \nu} = \eta \left(\Delta^\mu u^\nu + \Delta^\nu u^\mu - \frac{2}{3}
\Delta^{\mu \nu} \nabla_\alpha u^\alpha\right)\,,
\label{Smunu}
\end{eqnarray}
with the coefficient of shear viscosity $\eta$, and $\nabla^\mu =
\Delta^{\mu \nu} \partial_\nu$. The viscous stress-energy tensor $\pi^{\mu \nu}$ is clearly a complicated
object that is evaluated dynamically. Finally, vorticity and numerically
small terms have been neglected.

Hydrodynamic calculations require their initial states to be defined. In
this work, smooth or averaged initial conditions (AIC) will be considered,
as well as cases where these initial states are allowed to fluctuate (FIC)
about that average. The procedure for implementing AICs in \textsc{music}
is described in detail in Ref. \cite{Schenke:2010nt}, and that for FICs, in
Ref. \cite{Schenke:2011tv}; those descriptions will not be repeated here.
The initial time for the hydro to start is defined by $\tau_0$, in
this work this value is set to $\tau_0$ = 0.2 fm/c, and the freeze-out
energy density is 0.12 GeV/fm$^3$, which approximately corresponds to T =
137 MeV.

Considering first AICs, it is instructive to study how the inclusion of
shear viscosity affects the bulk evolution. The physical case being considered
is that
of Au + Au, at $\sqrt{s} = 200$ GeV, at an impact parameter of $b = 4.47$
fm, which represents a 0 - 20 \% centrality class. 
As mentioned earlier, the equation of state used here is
the parametrization ``s95p-v1'' from Ref. \cite{Huovinen:2009yb}. 
In this parametrization, the fit to the lattice QCD data is 
made above $T = 250\,\hbox{MeV}$ with the constraint that at
$T=800\,\hbox{MeV}$, the energy density reaches 95\,\% of the
Stefan-Boltzmann value. 
It is also worth noting that this parameterization correctly reproduces the trace anomaly
around the transition temperature: see Ref. \cite{Huovinen:2009yb} for more details.

\begin{figure}[h]
\begin{center}
{\includegraphics[width=8cm]{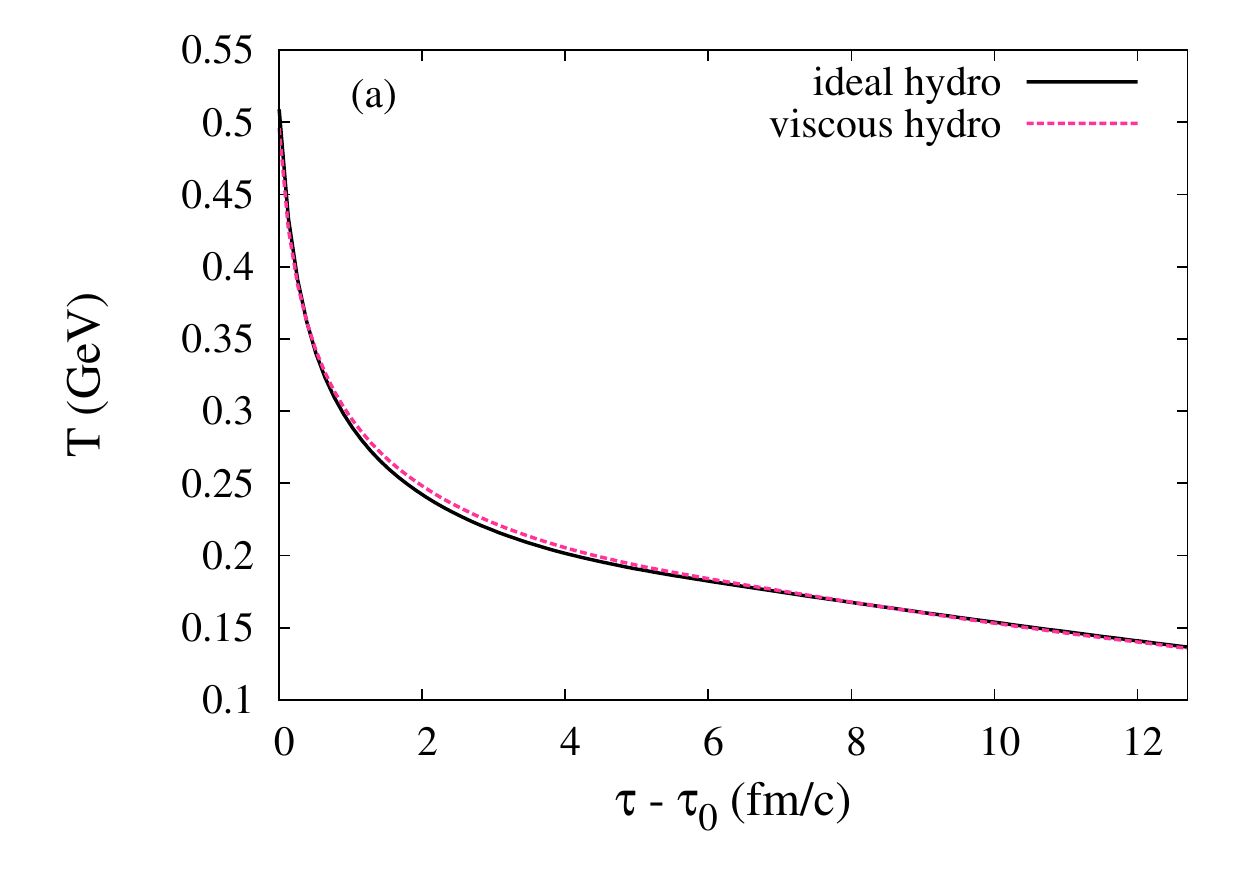}
\includegraphics[width=8cm]{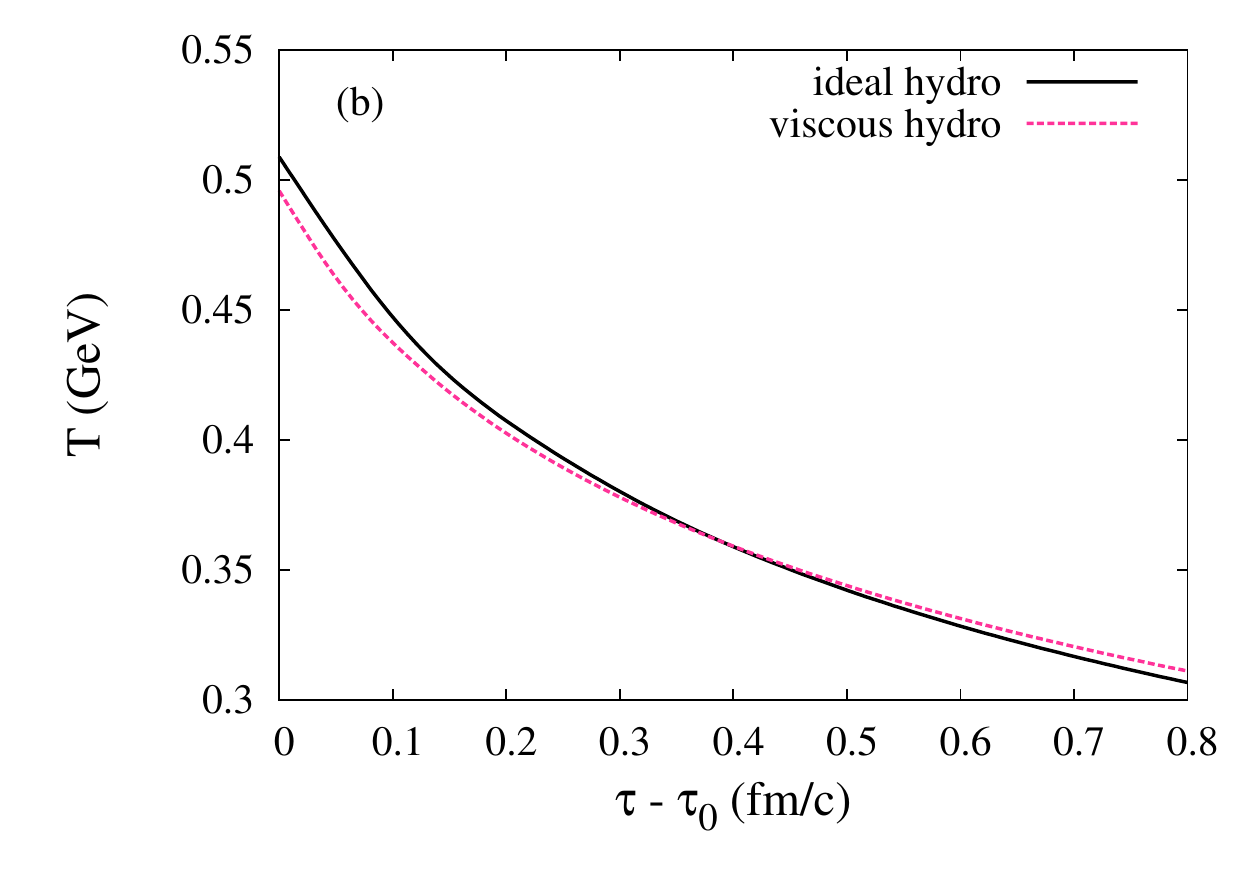}
\caption{(Color online) Left panel: The evolution of temperature in a fixed
cell, as a function of time. Results for ideal and viscous hydrodynamics
(with $\eta/s = 1/4 \pi$) are shown. Right panel: A closer view of the
early time temperature evolution.}
\label{T_evo}}
\end{center}
\end{figure}

Figure \ref{T_evo} shows the evolution of
temperature for a fixed cell at $x = y = 2.5$ fm and $z = 0$.
In the case with a non-zero shear viscosity, a value of $\eta/s$ (the shear
viscosity divided by the entropy density) = $1/4 \pi$ has been used. This
value has been suggested as a lower universal bound \cite{Kovtun:2004de}, a
statement which has raised some controversy and thus needed to be qualified 
\cite{Mia:2009wj,Buchel:2008vz,Cherman:2007fj}. Furthermore, $\eta/s$ will
also depend on the local temperature of the medium, but a constant value
will suffice for the study in this work. Note that the finite viscosity
calculation has a smaller initial temperature  than the ideal one, as
entropy will grow in the viscous case, affecting the observed final
particle multiplicity: it is important to compare calculations consistent with a given set of  hadronic observables.

\section{Photon emission from ideal and viscous media}
\label{viscous}

Viscous corrections on microscopic processes involving particles have been
included by writing the in-medium distribution functions with an
out-of-equilibrium correction, $f_0 \to f_0 + \delta f$, where $f_0$ is an
ideal Bose-Einstein/Fermi-Dirac distribution function. This is most easily
seen by considering the particle spectra being generated from the
Cooper-Frye formalism \cite{Cooper:1974mv}, and requiring that the energy
momentum tensor be continuous across the freeze-out hypersurface. In a
multispecies ensemble, a popular ansatz that satisfies the continuity
requirements is
\begin{eqnarray}
\delta f_i = f_{0 i} \left(1 \pm f_{0 i} \right) p^\alpha p^\beta \pi_{\alpha
\beta} \frac{1}{2 \left( \epsilon + P\right) T^2}
\label{visc_corr}
\end{eqnarray}
for the distribution function of species $i$. This form is used in this
work. In general, there can be an overall constant, different for each
species, that multiplies Eq. (\ref{visc_corr}) \cite{Dusling:2009df}. It is
implicit in this treatment that $\delta f$ should represent a small
correction. In the calculation of pions and of other hadronic observables,
one simply needs to verify this statement on the freeze-out hypersurface.
This is not the case for electromagnetic emission which occurs at all
time-scales of the hydro evolution. Therefore, photon calculations in
viscous media will represent a  stringent test of the validity of the
viscous dynamics, as shall be seen later.

\subsection{Photon emission from the QGP}

\begin{figure}[ht]
\includegraphics[width=4cm]{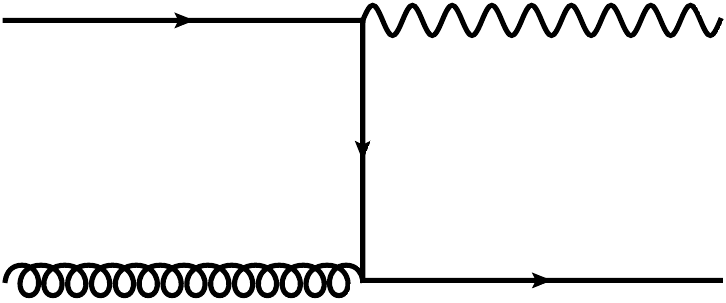}\hspace*{1.2cm}
\includegraphics[width=4cm]{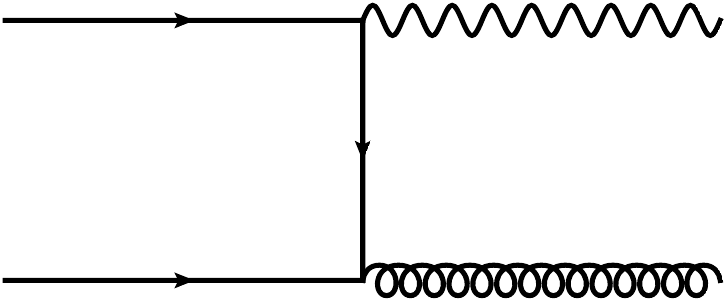}
\caption{The Compton and quark-antiquark annihilation
contributions to photon production.}
\label{Compt}
\end{figure}
\begin{figure}[ht]
\includegraphics[width=8.5cm]{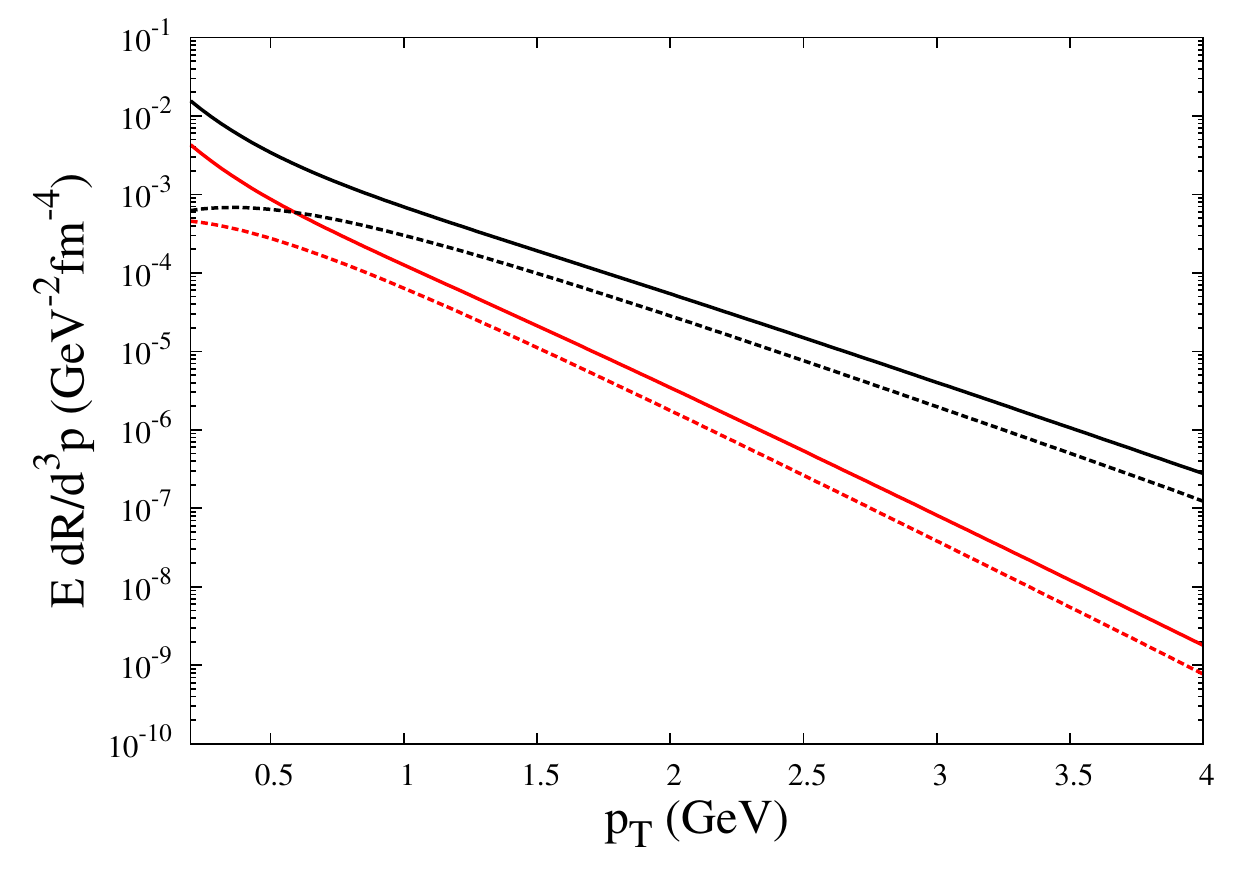}
\caption{(Color online) A comparison of the  equilibrium photon rate from the processes
shown in Figure \ref{Compt} (dashed lines) with that obtained tallying all
channels contributing at leading order in $\alpha_s$ (full lines), for $N_f
=3$. The lower set of curves are for T = 250 MeV, and the upper ones  are
for T = 350 MeV.}
\label{rate_comp}
\end{figure}

Rates, complete at leading order in $\alpha_s$, for the emission of photons
from a thermal ensemble of partons have now been available for a decade
\cite{Arnold:2001ms}. The extension of these results to viscous media
necessitates revisiting the resummation procedure in Ref.
\cite{Arnold:2001ms} with out-of-equilibrium distributions: a process we
shall not perform here. We rather concentrate on a subset of the diagrams: 
 the Compton and quark-antiquark annihilation processes shown in Figure
\ref{Compt}. It is instructive to compare the photon rate obtained through the approach
described above with the complete result at leading order in $\alpha_s$: this is done in Fig. \ref{rate_comp}. 
At low $p_T$, the full leading order rates are an order of magnitude
larger than the naive leading order rates owing to additional processes. For examples, the former receive a large contribution of
brehmsstrahlung from quarks of all momenta in this range. For $p_T >$ 1 GeV,
the full leading order rates are only larger by about a factor of two (there is however some temperature dependence to the position of this transition
window).

The net  photon emission rate $R$, summing these individual processes of the
type $1 + 2 \to 3 + \gamma$, is  obtained by evaluating
\begin{eqnarray}
E \frac{d^3 R}{d^3 p} &=& 
\sum_i
\frac{\cal N}{(2 \pi )^7} \frac{1}{16 E} 
\int ds dt | {\overline{{\cal M}_i}} |^2 \int d E_1 d E_2 f_1 (E_1) f_2 (E_2)
\nonumber \\ &&\hfill  \times \left[ 1 \pm f_3 (E_1 + E_2 -E)\right]
\frac{\theta \left(E_1 + E_2 - E \right)}{ \sqrt{\left(a E_1^2 + b E_1 + c
\right)}}
\label{phot_rate}
\end{eqnarray}
where the coefficients $a, b, c$ are defined in Eq. (\ref{abc_coeff}), and 
where $| \overline{{\cal M}_i} |^2 = 16 \pi s^2 d \sigma_i / dt$, with
\begin{eqnarray}
\frac{d \sigma_{\rm annihil.}}{d t} &=& \frac{8 \pi \alpha \alpha_s}{9
s^2}\frac{u^2 + t^2}{u t}\ ,\ \nonumber\\
\frac{d \sigma_{\rm Compt.}}{d t} &=& \frac{ - \pi \alpha \alpha_s}{3 s^2}
\frac{u^2 + s^2}{u s}
\end{eqnarray}
Note also the degeneracy factors ${\cal N}_{\rm annihil.}= 20$, and ${\cal
N}_{\rm Compt.} = 320/3$, for $N_f = 2$. In the case of $N_f =3$, those
numbers become ${\cal N}_{\rm annihil.}= 24$, and ${\cal N}_{\rm Compt.} =
384/3$.
Calculations of the photon production rate from these channels were done in
Ref. \cite{Kapusta:1991qp}, an evaluation with general anisotropic distribution functions (not limited to small deviations from equilibrium) appeared in Ref. \cite{Schenke:2006yp},  and a viscosity-corrected rate (to first order in $\delta f$) 
was obtained
recently in \cite{Dusling:2009bc}, assuming  forward-scattering dominance of the photon-producing reaction.  The rates reported here are obtained through a
numerical integration of Eq. (\ref{phot_rate}) with out-of-equilibrium distribution functions (Eq. (\ref{visc_corr})). The integrations span the entire accessible phase space, carefully avoiding
divergences as prescribed in Ref. \cite{Kapusta:1991qp}. Appropriate quantum statistics have been used.

\subsection{Photon emission from the hadronic gas}
As the ensemble of partons thermalizes (totally or partially) and then
expands and cools, it hadronizes into an ensemble of colorless
hadrons called here the hadronic gas (HG) which continues to expand and
to cool even more. The HG  thermal electromagnetic emissivity has been
characterized in Ref. \cite{Turbide:2003si}. Following that reference, a Massive Yang-Mills (MYM) model is used to model the
interactions between light pseudoscalars, vector and axial vector mesons.
The set we consider contains the elements $\{ \pi, K, \rho, K^*, a_1 \}$,
and the most important photon-producing rates are $\pi + \rho \to \pi +
\gamma$, $\pi+\pi \to \rho + \gamma$, $\pi + K^* \to  K + \gamma$, $\pi + K
\to K^* + \gamma$, $\rho + K \to  K + \gamma$, $K^* + K \to \pi + \gamma$. Two-body photon-production processes dominate the phase space for photon transverse momenta above 0.5 GeV \cite{Turbide:2003si}. 
All
isospin-allowed channels are considered. 

The viscous
corrections also demand a complete recalculation of the HG photon rates, by
including the corrected distribution functions - see Eq. (\ref{visc_corr})
- in all the relevant rate equations. Note that corrections of order
$\delta f^2$ are neglected for consistency, as are corrections to
Pauli-blocking or Bose-enhancement effects. These corrections are found to
be small.  The Appendix outlines the procedure for correcting the
electromagnetic emissivities, allowing for viscous effects in the hadronic
distribution functions.

\section{Results}
\label{results}

\subsection{Viscous corrections: generalities}

For both cases discussed in the previous section (QGP and HG), rates for
``viscous photons'' were not shown. In fact, those require detailed
dynamical information as they depend on the details of $\pi^{\mu \nu}$ and of
its time evolution as specified by Eqs. (\ref{visc_corr}) and
(\ref{W_evol}).
\begin{figure}[ht]
\includegraphics[width=8.5cm]{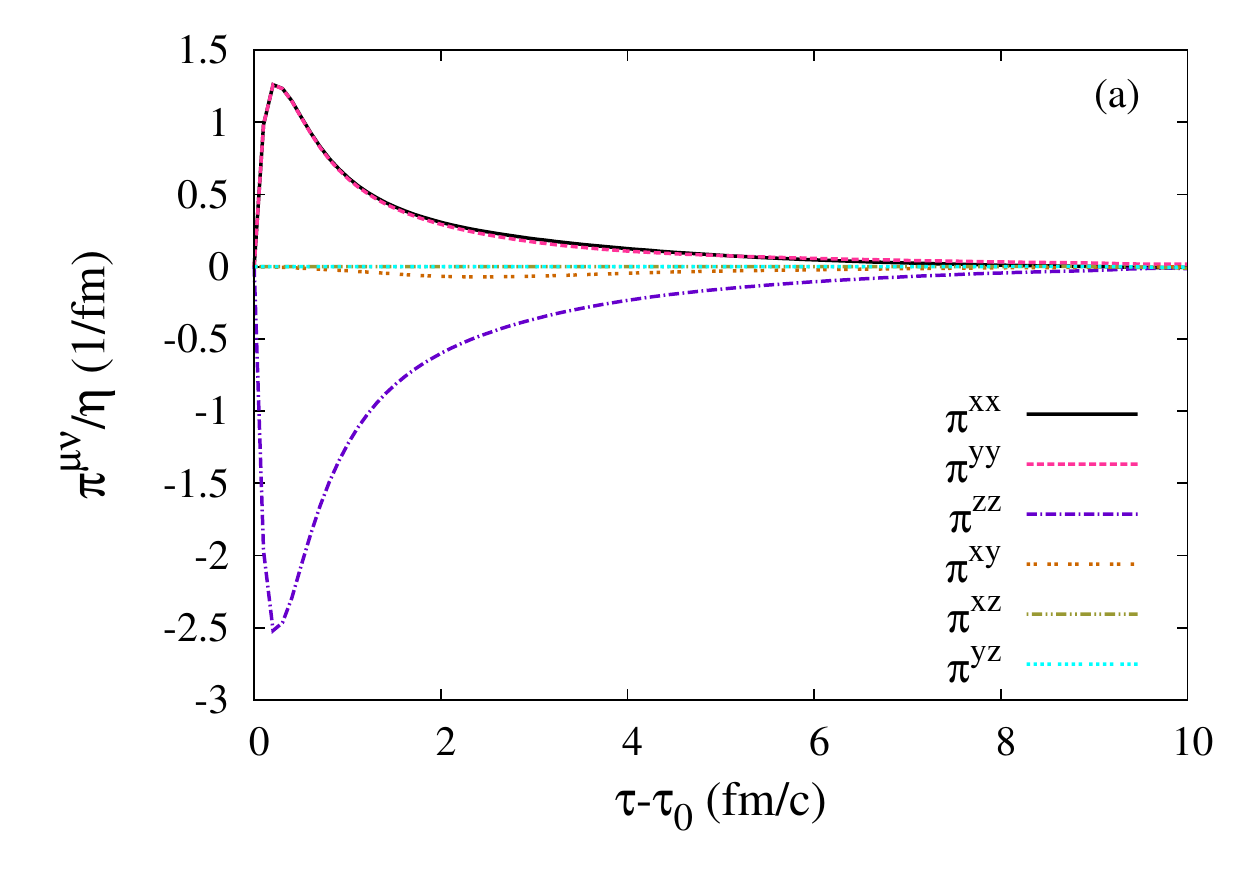}
\includegraphics[width=8.5cm]{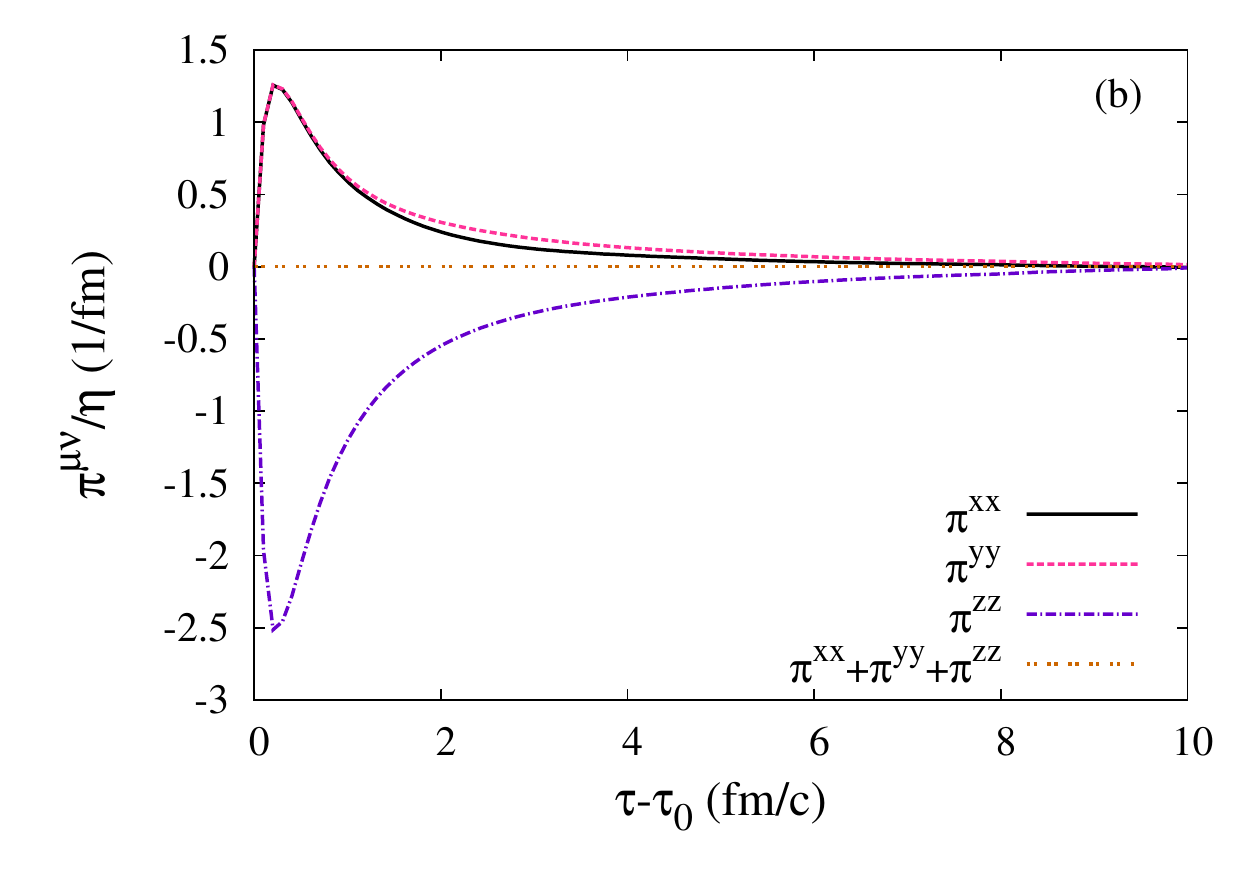}
 \caption{(Color online) Left panel: The time evolution of  different components of the
local $\pi^{\mu \nu}$ tensor, divided by $\eta$. Right panel:  The time evolution of the diagonal elements of $\pi^{i j}$ (scaled by $\eta$), and also that of the trace of the viscous tensor. The calculations are done for a fluid cell
at $x = y =$ 2.5 fm, and $z = 0$, and the impact parameter is $b = 4.47$ fm.}
\label{Wmunu}
\end{figure}
It is thus appropriate to examine this quantity here, and
this is done in Figure \ref{Wmunu}, in the rest frame of a fluid cell; note
that there $\pi^{tt}$ is 0. { At the initial time, the viscous corrections are non-existent, as we initialize the viscous pressure tensor to zero. They build up quickly, and then decay back to zero. Right after the initial time, }the magnitude of the
$zz$ component is larger than the other two diagonal ones
by roughly a factor of 2, and this fact persists up { to} late times.
The relative sign of $\pi_{zz}$ can be understood from the fact that
$\pi_{ij}$ should be traceless in the fluid rest frame (c.f. Eqs. (\ref{W_evol}, \ref{Smunu})). Note that this requirement was not enforced explicitly at each step of the calculation. The preservation of this trace then reflects the stability of the numerics: see the right panel of Figure \ref{Wmunu}.   
The slight difference between $\pi_{xx}$ and $\pi_{yy}$ is to be expected because of the elliptic
shape of the system: the $x-y$ symmetry is broken by the finite impact parameter.

\begin{figure}[h]
\includegraphics[width=8.5cm]{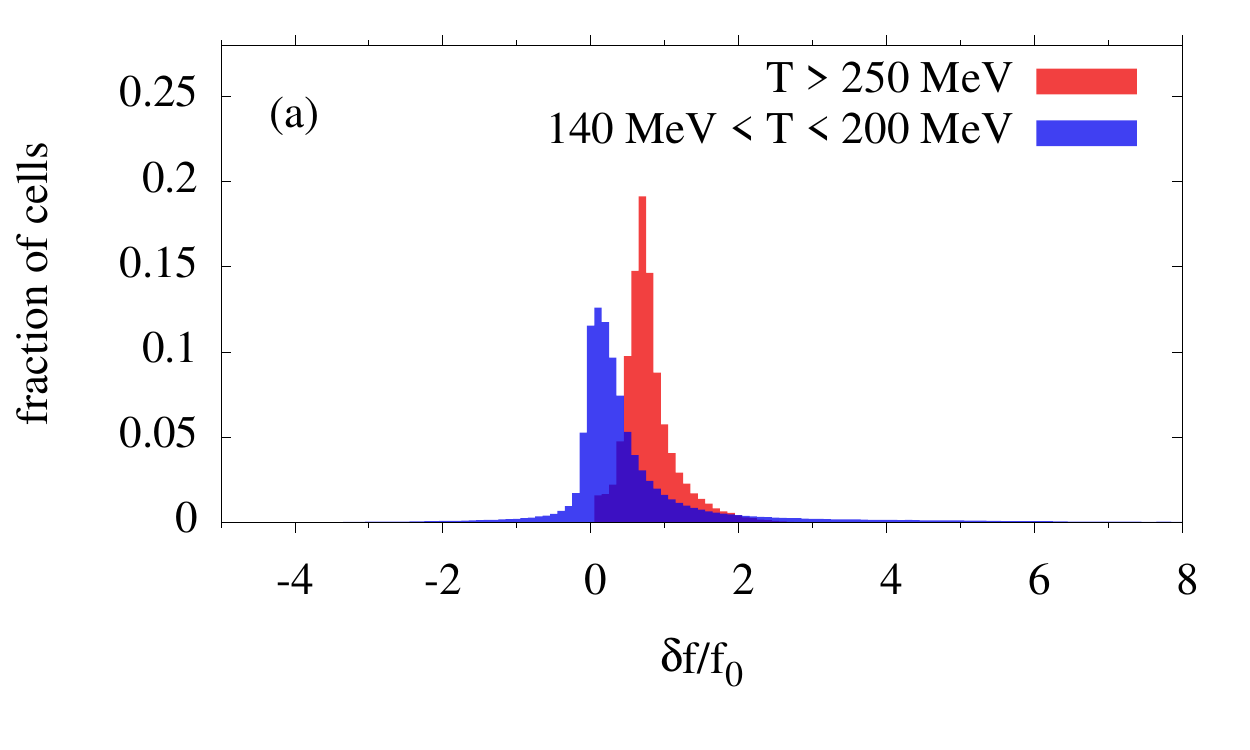}
\includegraphics[width=8.5cm]{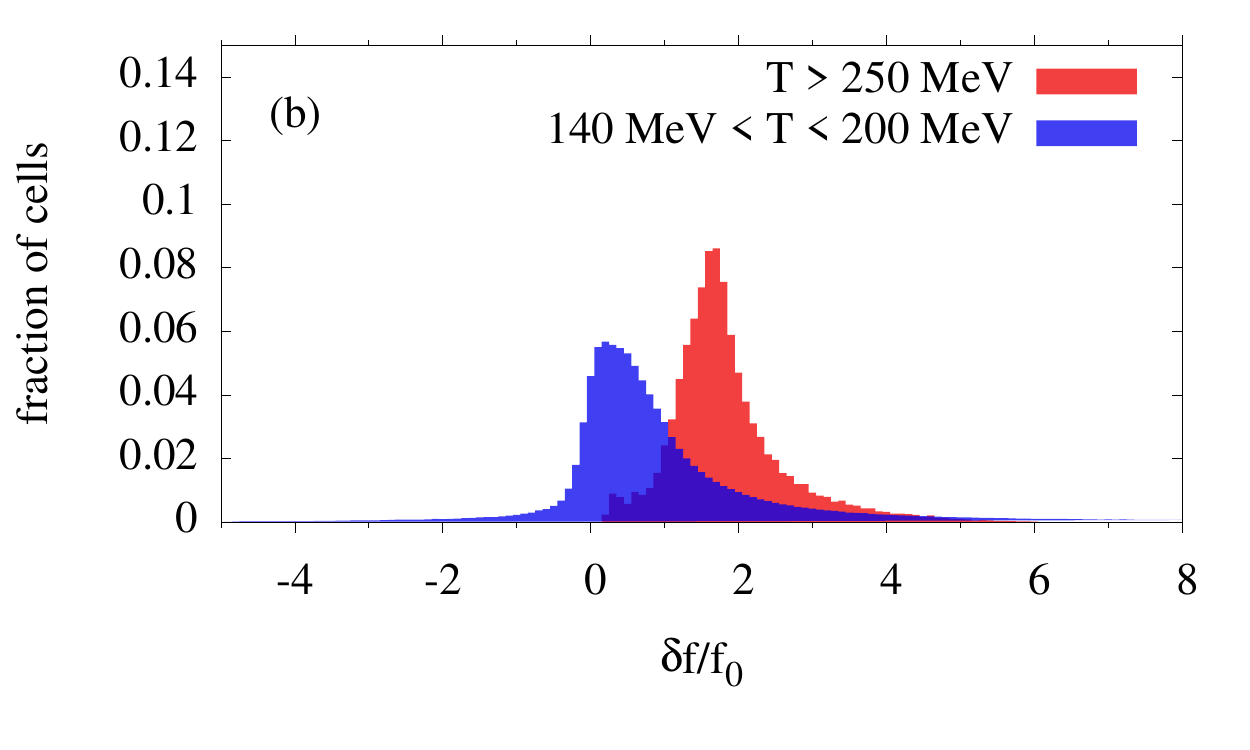}
\caption{(Color online) The fraction of $\eta_s \approx 0$ fluid cells with
a certain value of $\delta f / f_0$, for different values of the photon
momentum in the nucleus-nucleus  center of mass 
frame: 2 GeV (left panel), and 3 GeV (right
panel). The range of temperature $T > 250$ MeV corresponds to $\tau -
\tau_0 \lesssim$ 2 fm/c.}
\label{histo_1}
\end{figure}
To get a qualitative picture and develop some intuition for the importance
of the viscous corrections, the following procedure was implemented. In the
nucleus-nucleus center of mass frame, one picks a photon  momentum in the $x-y$ plane, at an angle
of $\pi/4$. The $z$ axis is the beam axis. Lorentz-transforming to the rest-frame of each fluid cell, the local value of the photon momentum is
obtained. Since  photons are formed in 2 $\to$ 2 processes, 
the magnitude of
this momentum is then roughly equal to the magnitude of the momentum of one of the interacting
particles. 
Finally, posing that this particle is a massless fermion will
enable a determination of the viscous correction to its distribution
function. This study is restricted to a slice in space-time rapidity,
$\eta_s$, centered around 0. This procedure is clearly approximate - not
all particles are massless and not all are fermions - but should
nevertheless produce a result indicative of the physics at play. Defining
bins of size $0.1$ in the relative variable $\delta f / f_0$, the fraction of $\eta_s
\approx 0$ cells with a certain value of this relative variable is plotted
in Figure \ref{histo_1}. In addition, in this study we concentrate on two ranges
of temperature: one corresponding to ``early times'', and another
corresponding to ``late times'' according to Figure \ref{T_evo}. The photon
energies chosen are typical values of the photon spectrum, see the next
subsection.  For a photon energy of 2 GeV, one sees that 
 $\approx 20\%$
of the fluid cells have a $\delta f / f_0 \geq 1$, at early times, 
and that the distribution around this value is fairly narrow.
For a higher
photon energy of 3 GeV, this distribution has grown in width, now { with 80\% of the high-temperature cells with $\delta f/f_0 \geq 1$} and 
$\approx$ 30\% of them with $\delta f / f_0 \geq 2$: a clear violation of the perturbative nature of the approximation.  In a given panel of Figure \ref{histo_1}, the amount of larger viscous corrections at higher
temperatures can be understood: those cases correspond to situations at early
times where the elements of the $\pi^{\mu \nu}$ tensor are large (see Figure
(\ref{Wmunu})). Higher momenta will command larger viscous
corrections, see Eq. (\ref{visc_corr}), and hence the broadening of the distributions for a given
range in $T$ when going from the left to the right panel of Figure
\ref{histo_1}. Note that repeating the lower temperature part of this
analysis by assuming that the corrected particles are massive $\pi$'s or even
massive $\rho$'s do not change its conclusions. { A negligible but non-zero number of high-momentum cells have $\delta f/f_0 \leq -1$. For those cells the photon emission probability has been set to zero. }What emerges here is an explicit 
message of caution. In contrast to hadron calculations where only
properties at the freeze-out surface are required, the evaluation of
electromagnetic signals requires the {\it entire} time evolution to be
monitored. In the case of the analysis shown here, one should also keep in mind that thermal photons with an energy equal to, or greater than, 3 GeV will  lie below those from other sources like the direct photons from primordial nucleon-nucleon collisions, for example \cite{Turbide:2007mi,Qin:2009bk}. 

One  now proceeds to the evaluation of
photon characteristics, and of the influence of viscous effects on them. All
of the calculation results shown in the next sections will rely on the use
of Averaged Initial Conditions (AICs). A discussion of the effects of
Fluctuating Initial Conditions (FICs) on real photons first appears in
section \ref{FIC}.

\subsection{Photon spectrum}

\begin{figure}[h]
\includegraphics[width=8cm]{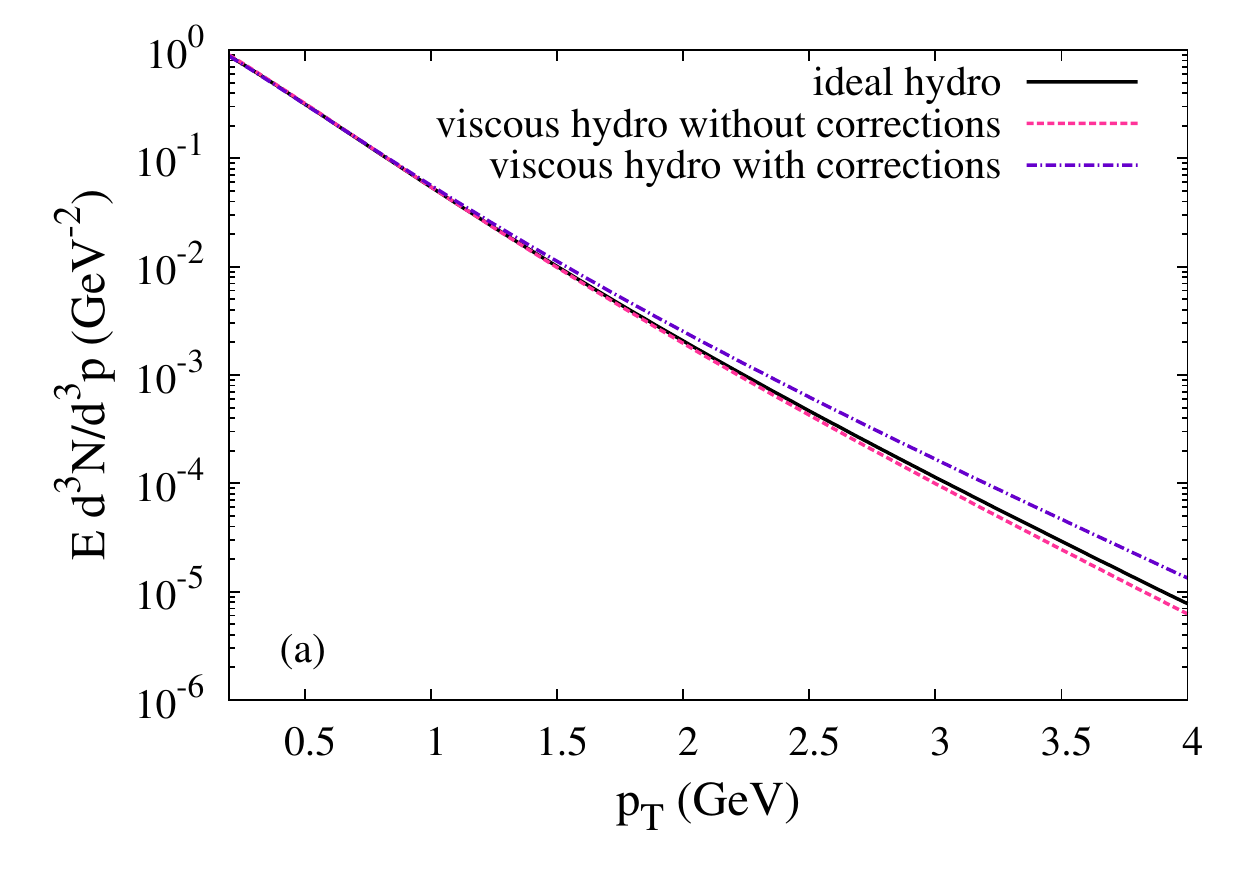}
\includegraphics[width=8cm]{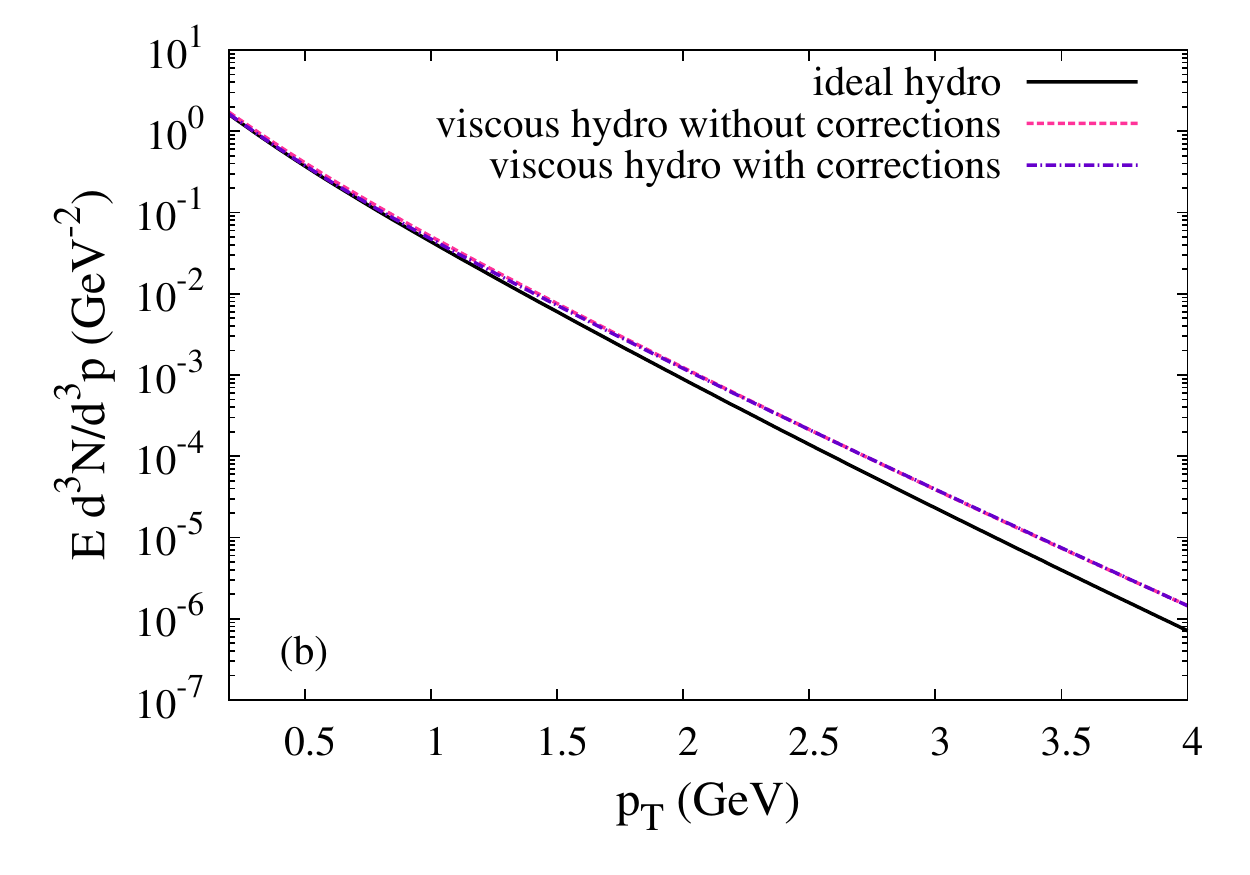}
\caption{(Color online) Left panel: The photon yield originating from the phase with
parton degrees of freedom only. The contribution from ideal hydro is shown
(solid  curve), together with the result of using a time evolution
associated with viscous hydrodynamics (dotted line), and using a viscous
time evolution and corrected  microscopic distribution functions
(dash-dotted line). Right panel: The photon yield originating from the hadronic gas
only. The meaning of the different curves is the same as that in the left panel.}
\label{QGP_yield}
\end{figure}
As mentioned previously, the measured photon spectrum receives
contributions from all times and all phases spanned by the collision
dynamics. There are treated in turn here, for pedagogical purposes.
The photons originating solely from the QGP phase are first shown in Figure
\ref{QGP_yield}, and they are obtained by integrating the rate in Eqs.
(\ref{phot_rate}) and (\ref{visc_rate}) throughout the time
evolution dictated by \textsc{music}. 

As compared to an ideal hydrodynamical evolution, the viscous evolution starts with a lower initial temperature when the system is entirely in its QGP phase (see Figure \ref{T_evo}). Therefore, integrating the QGP
photon rates with a viscous hydrodynamic evolution  alone 
produces a photon
spectrum slightly lower at high values of $p_T$ than that generated by the
ideal hydro. This is the dotted line in Figure \ref{QGP_yield}. Then, using the corrected distribution functions
make the photon spectrum harder, as the correction grows as a function of
momentum. This is the upper curve in the same figure. The hardening of the QGP photon spectra owing to shear viscous effects had also been noticed in previous work \cite{Dusling:2009bc,Bhatt:2009zg,*Bhatt:2010cy,Chaudhuri:2011up}. 

\begin{figure}[h]
\includegraphics[width=8cm]{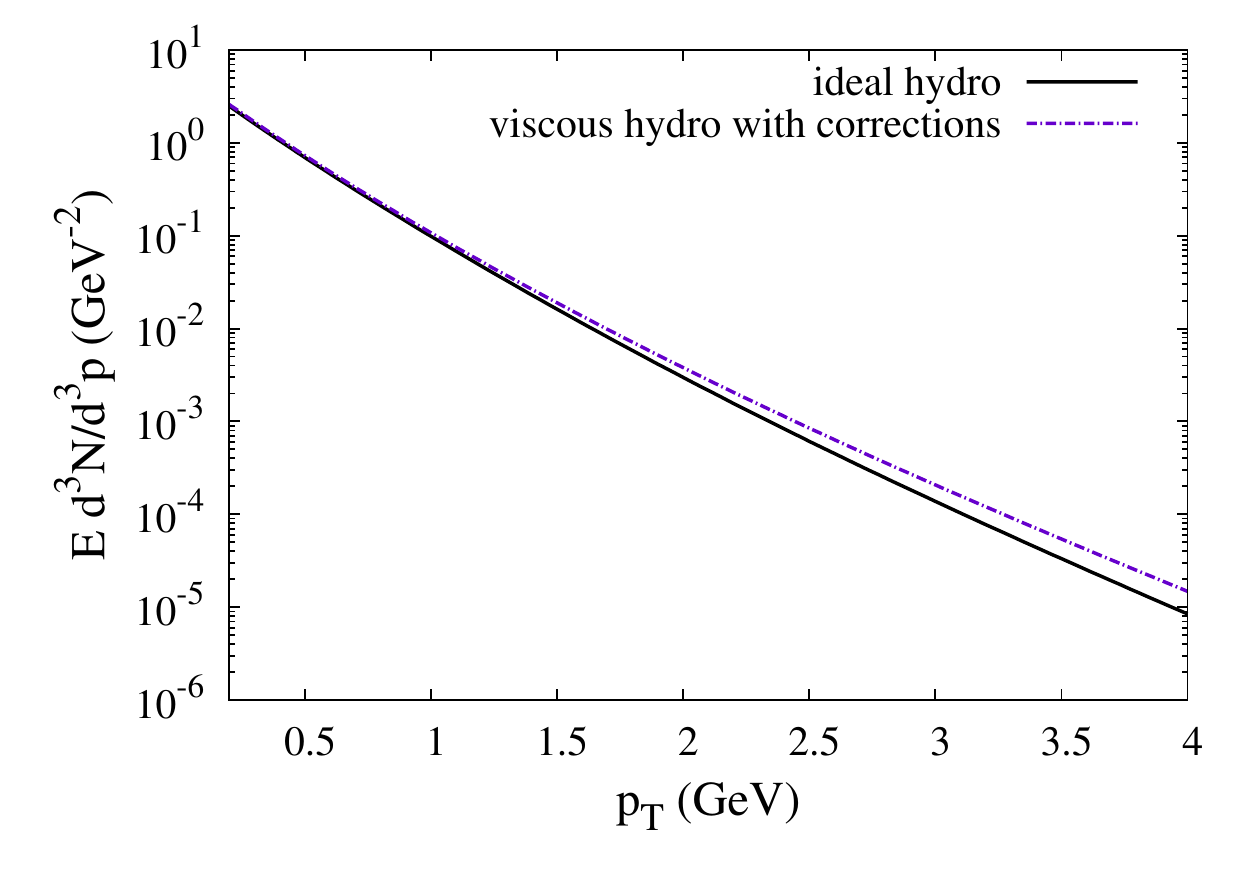}
\caption{(Color online) The net thermal photon yield, from QGP and HG
sources. The ideal spectrum (i.e. using an ideal hydrodynamics background),
and the viscous spectrum (using a viscous hydrodynamics background and
corrected microscopic distribution functions) are shown as a solid and
dotted line, respectively.}
\label{total_yield}
\end{figure}

Turning now to photons originating solely from the HG sector, the relevant
spectrum is shown in the right panel of Figure \ref{QGP_yield}. Interestingly, the spectrum
with the viscous corrections (viscous hydro and corrected distribution
functions) is essentially undistinguishable from that obtained using ideal
rates integrated with a viscous time evolution.  This can be understood by considering the fact that photons from the HG are emitted later in time, essentially when $\pi^{\mu \nu} \sim 0$, as is made clear in Figure \ref{Wmunu}.  The effect of viscosity are manifested in a slightly harder spectrum: in part a consequence of the temperature in the viscous evolution remaining  higher than that in
the ideal evolution for intermediate and late times, as shown in
Figure \ref{T_evo}. The yield of real photons from all thermal sources (QGP + HG) is shown in
Figure \ref{total_yield}, for an ideal hydrodynamic evolution and also for
a viscous evolution (viscous hydro and corrected distribution functions).
The difference between the two scenarios is actually small at intermediate values of the photon transverse momentum, growing to being approximately 100\% at $p_T = 4$ GeV. At that energy however, the purely thermal
photons will lie below other sources and will be sub-dominant, as mentioned already. 
One may thus conclude here that
extracting information about the shear viscosity from photon spectra alone
will be an arduous task. More work is needed however to include {\it all} the
photon sources in a theoretically consistent way, with all the viscous
corrections.

\subsection{Photon elliptic flow}
\begin{figure}[h]
\includegraphics[width=8cm]{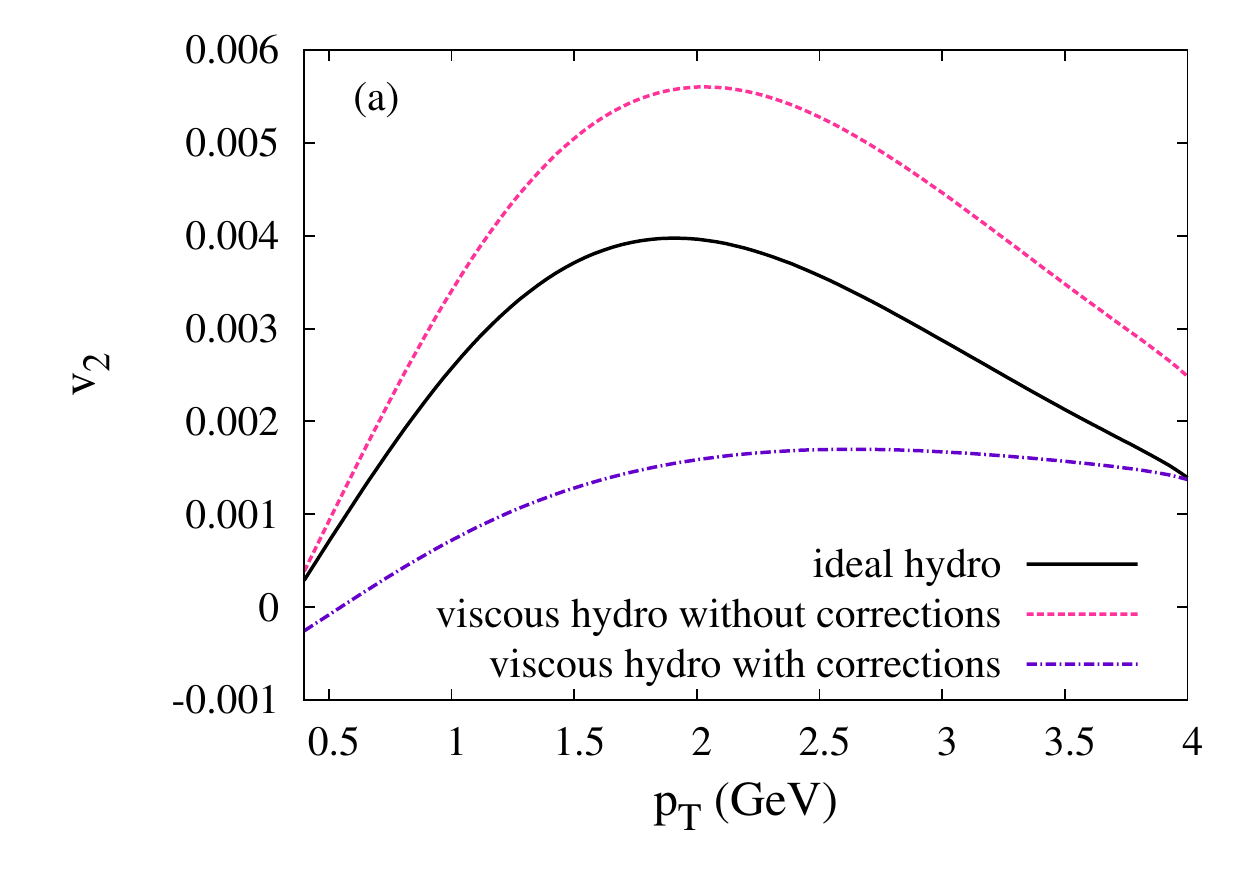}
\includegraphics[width=8cm]{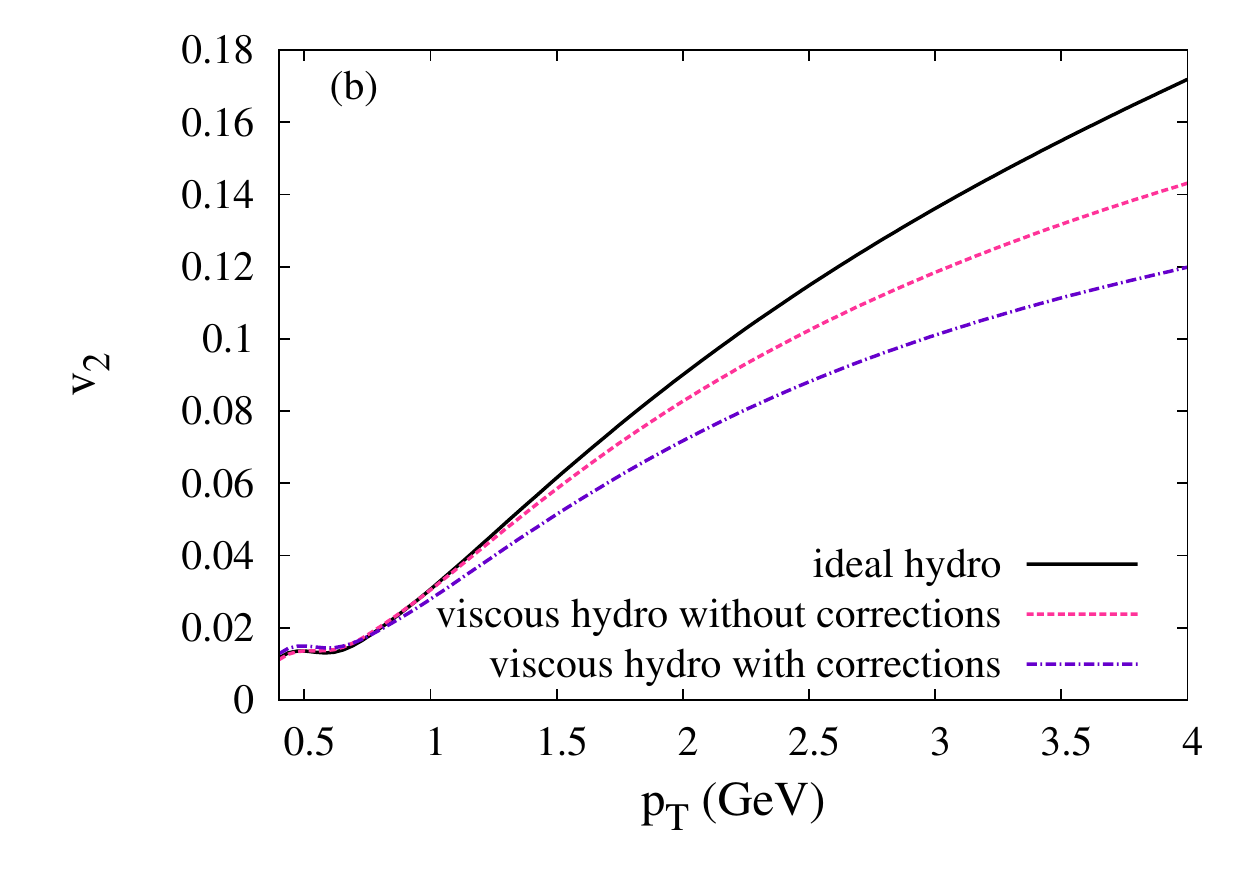}
\caption{(Color online) Left panel: The thermal photon elliptic flow, considering only
the photons originating from the QGP.  As in previous figures, the results
of using ideal hydrodynamics (solid line), viscous hydrodynamics with
equilibrium rates (dotted line), and viscous hydrodynamics with $\delta f$
corrections (dash-dotted line) are shown separately. Right panel: The thermal photon elliptic flow, considering only
the photons originating from the HG. The lines have the same meaning as those in the left panel.}
\label{QGP_v2}
\end{figure}

\begin{figure}[h]
\includegraphics[width=8cm]{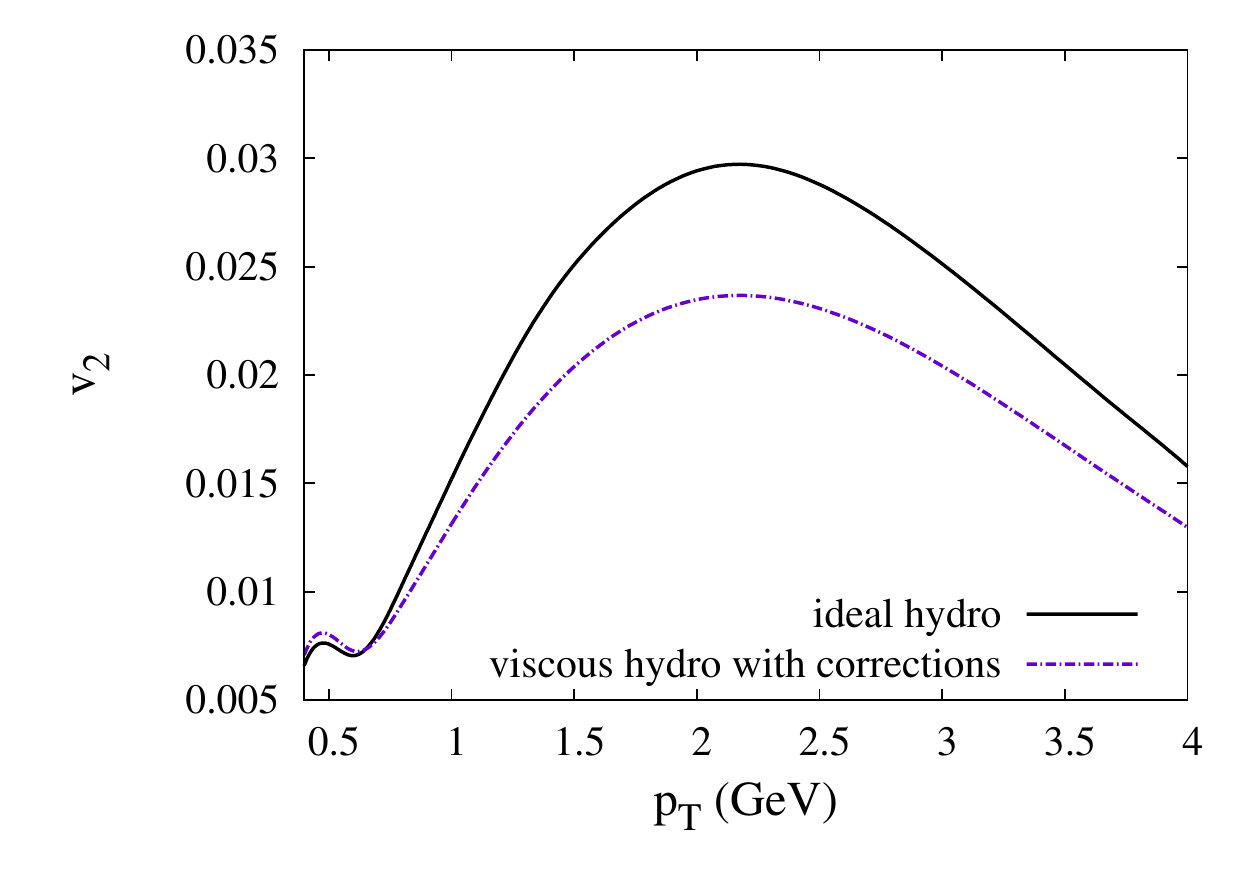}
\caption{(Color online) The net thermal photon elliptic flow. The curves
have the same meaning as in Figure \ref{total_yield}.}
\label{total_v2}
\end{figure}

The flow characteristics of hadrons have contributed considerably to
quantify the details of the underlying hydrodynamics, and this fact has
been hailed as one of the major milestones of the RHIC program. As for photons, their
elliptic flow holds the potential of providing more insight into the dynamics
of heavy ion collisions, and into the phase structure of QCD. Indeed, the
shape of the real photon $v_2$ coefficient is directly sensitive to the nature of
the underlying degrees of freedom \cite{Chatterjee:2005de,*Heinz:2006qy},
unlike the single-photon spectra of the previous section. The elliptic flow
of photons originating solely from the QGP is shown in Figure \ref{QGP_v2}.
If one neglects the correction to the distribution functions, the elliptic
flow from the viscous evolution appears slightly larger than in the ideal
case, reflecting an increase in the azimuthal asymmetry of the fluid flow
pattern due to viscosity, consistent with the large gradients at early
times implied by Figure \ref{Wmunu}.  However, the corrections to the
distribution functions dominate, and make the net anisotropy even smaller
than in the ideal case. { This behaviour is consistent with the results of Ref. \cite{Chaudhuri:2011up}.
It is also worthwhile to point out that the apparently negative values of $v_2$ at very low momenta for photons in the QGP are obtained only using the photon-production rates corresponding to Figure \ref{Compt}. The complete rates of Ref. \cite{Arnold:2001ms} do not yield negative photon elliptic flow in ideal hydrodynamics. The negative values also appear in earlier calculations \cite{Dusling:2009bc,Chaudhuri:2011up}.}
The HG $v_2$ is shown in the right panel of Figure  \ref{QGP_v2} and there, all viscous
corrections make the elliptic flow smaller, unlike the case for the QGP.
This is again a reflection of the richness of the dynamics contained in the
time-dependence of $\pi^{\mu \nu}$. Further note that the small structure at
low momenta signals a crossover between two different hadronic channels
\cite{Chatterjee:2005de,*Heinz:2006qy}. The net photon $v_2$ is then
calculated and shown in Figure \ref{total_v2}.
Importantly, the total $v_2$ is a weighted average of the individual (QGP, and HG)
coefficients, the weight being the value of the appropriate single-photon
distribution. Hence, in the computation of the final $v_2$, the small QGP $v_2$ will get multiplied by a large emission rate, whereas the smaller emission rate of the HG phase
gets partially compensated by the larger flows. Both phases therefore
contribute to the final profiles shown in Figure \ref{total_v2}.

\subsection{Fluctuating initial conditions (FIC)}
\label{FIC}
The recent years have witnessed  a paradigm-shift in the analysis of heavy
ion collision data. Up until recently, smooth initial state distributions
were mostly used  in hydrodynamics analyses of relativistic nuclear collisions.
These, together with conservation laws, imply that odd-numbered expansion
coefficients in Eq. (\ref{v_n}) vanish identically. As discussed in the
Introduction, this situation has changed  with the work of Ref.
\cite{Alver:2010gr} linking odd-numbered flow harmonics to initial state fluctuations.
The hydrodynamic simulation \textsc{music} with viscous corrections has
recently been modified to include FICs \cite{Schenke:2010rr}. This has been used to make a prediction for size and momentum dependence of  the hadronic $v_3$ at RHIC. This prediction has been recently confirmed \cite{Adare:2011tg}. Here we seek to assess the importance of the event-by-event fluctuations on photon
observables. 

For initial conditions that are not smooth, it is important to specify how the reaction plane is determined. The ``participant plane'' \cite{Holopainen:2010gz} is used here. Namely, one calculates event-by-event the angle $\psi_2$ with respect to the reaction plane defined by the impact parameter:
\begin{eqnarray}
\psi_2 = \frac{1}{2} \arctan \left( \frac{\langle r^2 \sin(2 \phi) \rangle}{\langle r^2 \cos(2 \phi) \rangle }\right)
\end{eqnarray}
where the averages are over wounded nucleon positions, $(r, \phi)$, in the transverse plane. The angle $\psi_2$ then goes into the evaluation of $v_2$, with $\psi_2$ replacing $\psi_r$ in Eq. (\ref{v_n}). Note that the initial eccentricity is maximized by the choice of this participant plane. The studies performed here used ensembles of 50 events, leading to uncertainties of the order of 5\% on thermal photon spectra, and of the order of 15\% on thermal photon $v_2$. The precise value of these variations is of course $p_T$-dependent, but we find that elliptic flow does depend more strongly on the initial structure of the energy density distribution than the momentum spectrum.

\begin{figure}[h]
\includegraphics[width=8.5cm]{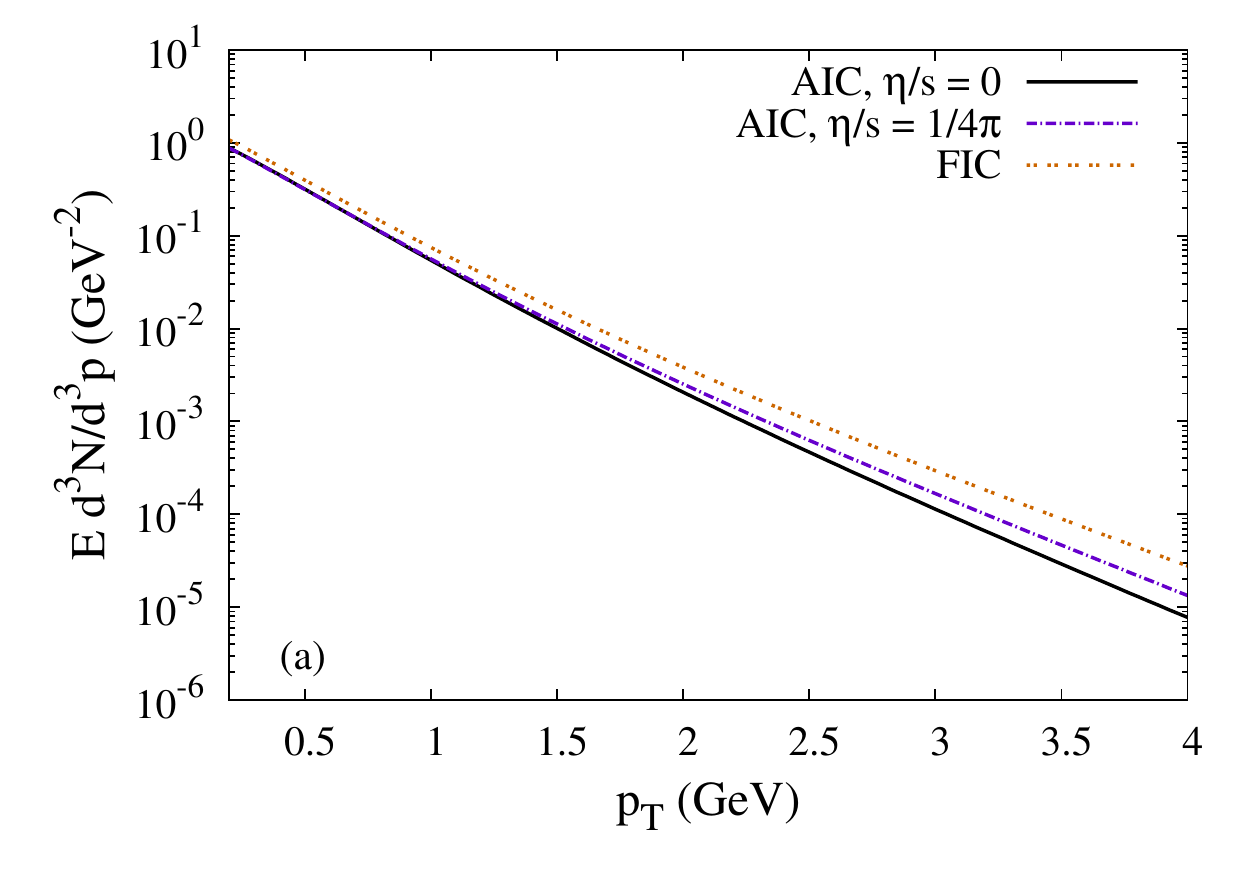}
\includegraphics[width=8.5cm]{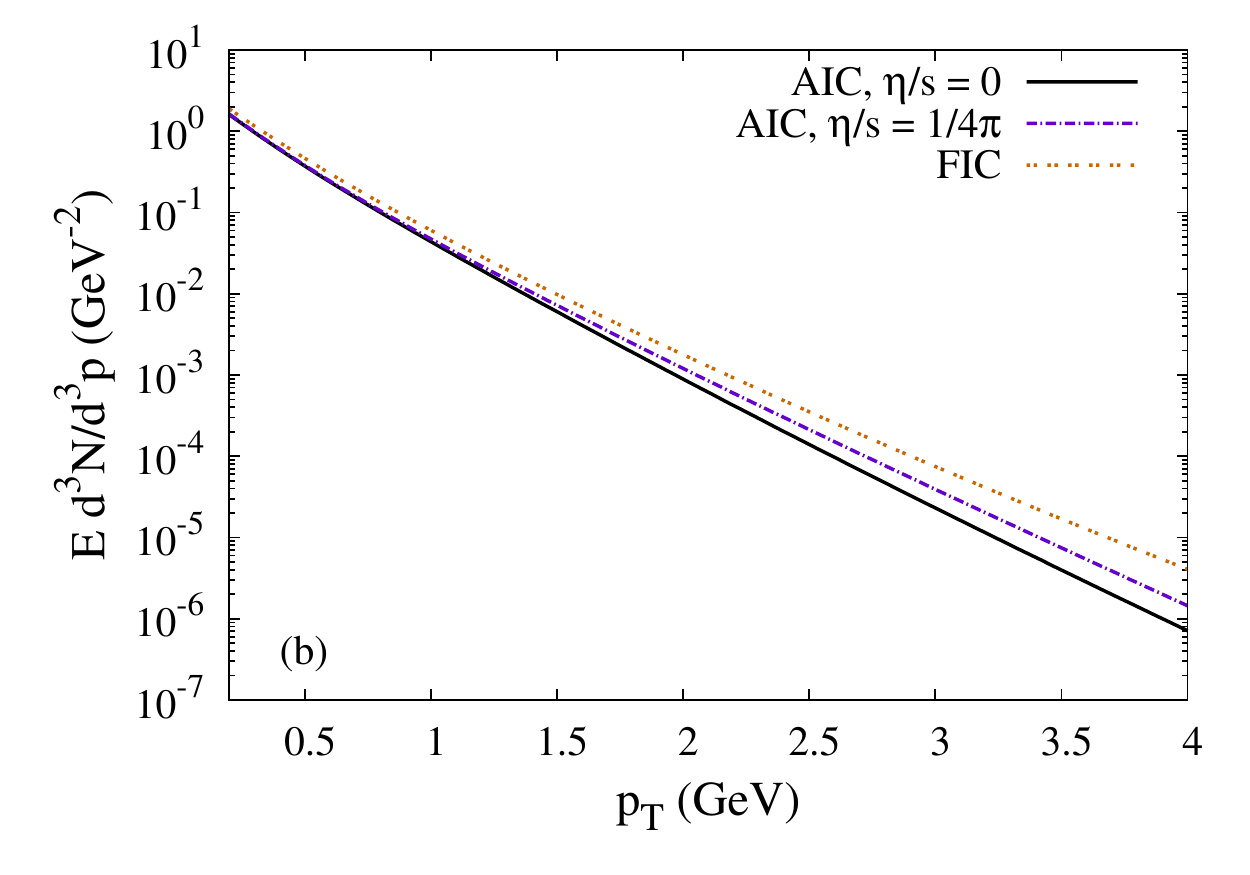}
\caption{(Color online) The thermal photon yield, showing the effect of
FICs. The left panel shows the contribution from the QGP, the right panel
that of the HG. Note that the curve labeled ``FIC'' also includes all
viscous corrections (time evolution and $\delta f$)}
\label{yield_FIC}
\end{figure}

As already observed for hadrons \cite{Qiu:2011iv} and more recently for photons
\cite{Chatterjee:2011dw}, the lumpy initial states lead to a yield enhancement.  Again, the QGP and HG contributions are calculated 
separately. They are shown in the two panels of Figure \ref{yield_FIC}, and the quantitative importance of the enhancement can be judged there.
\begin{figure}[h]
\includegraphics[width=8.5cm]{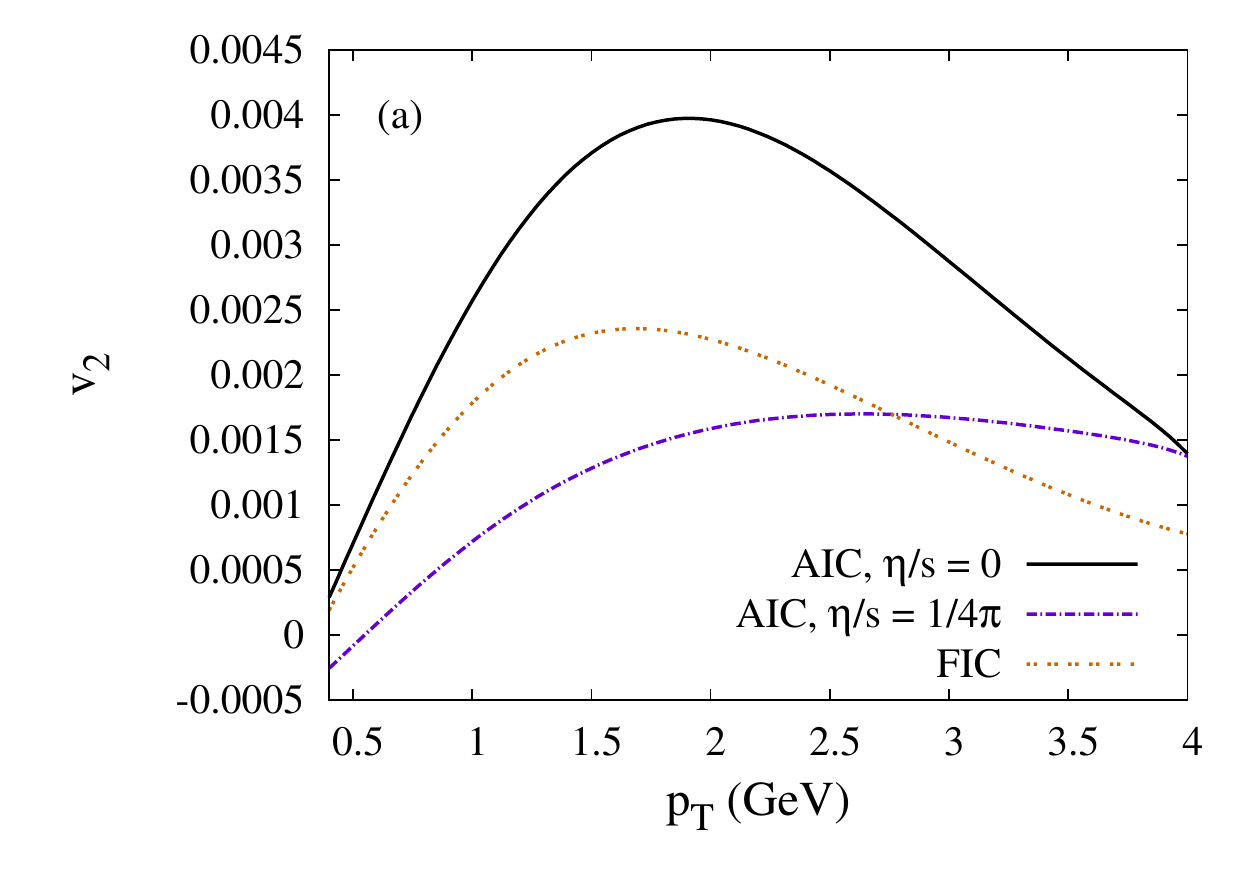}
\includegraphics[width=8.5cm]{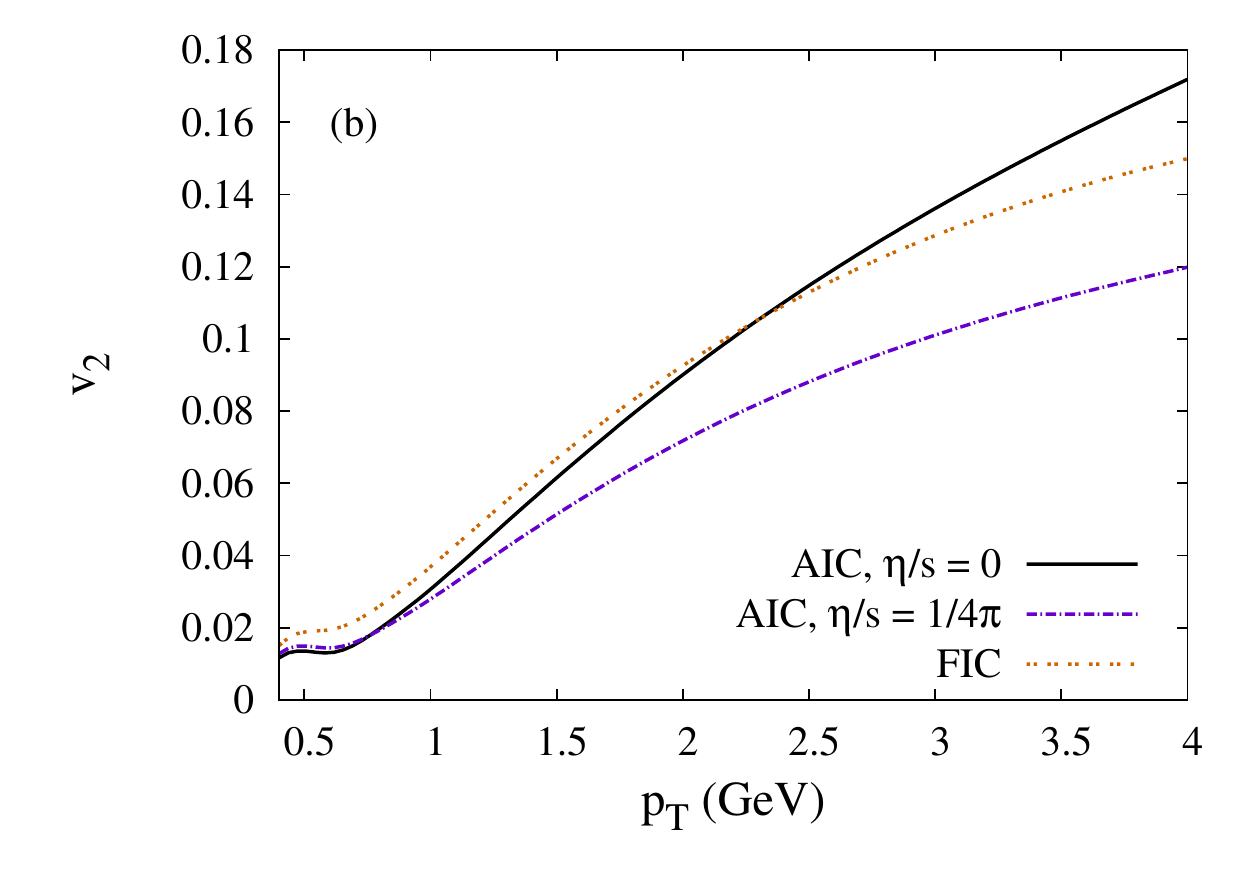}
\caption{(Color online) The thermal photon $v_2$, showing the effect of
FICs. The left panel shows the contribution from the QGP, and the right panel
that of the HG. Note that the curve labeled ``FIC'' also includes all
viscous corrections (time evolution and $\delta f$).}
\label{v2_FIC}
\end{figure}
\begin{figure}[h]
\includegraphics[width=8.5cm]{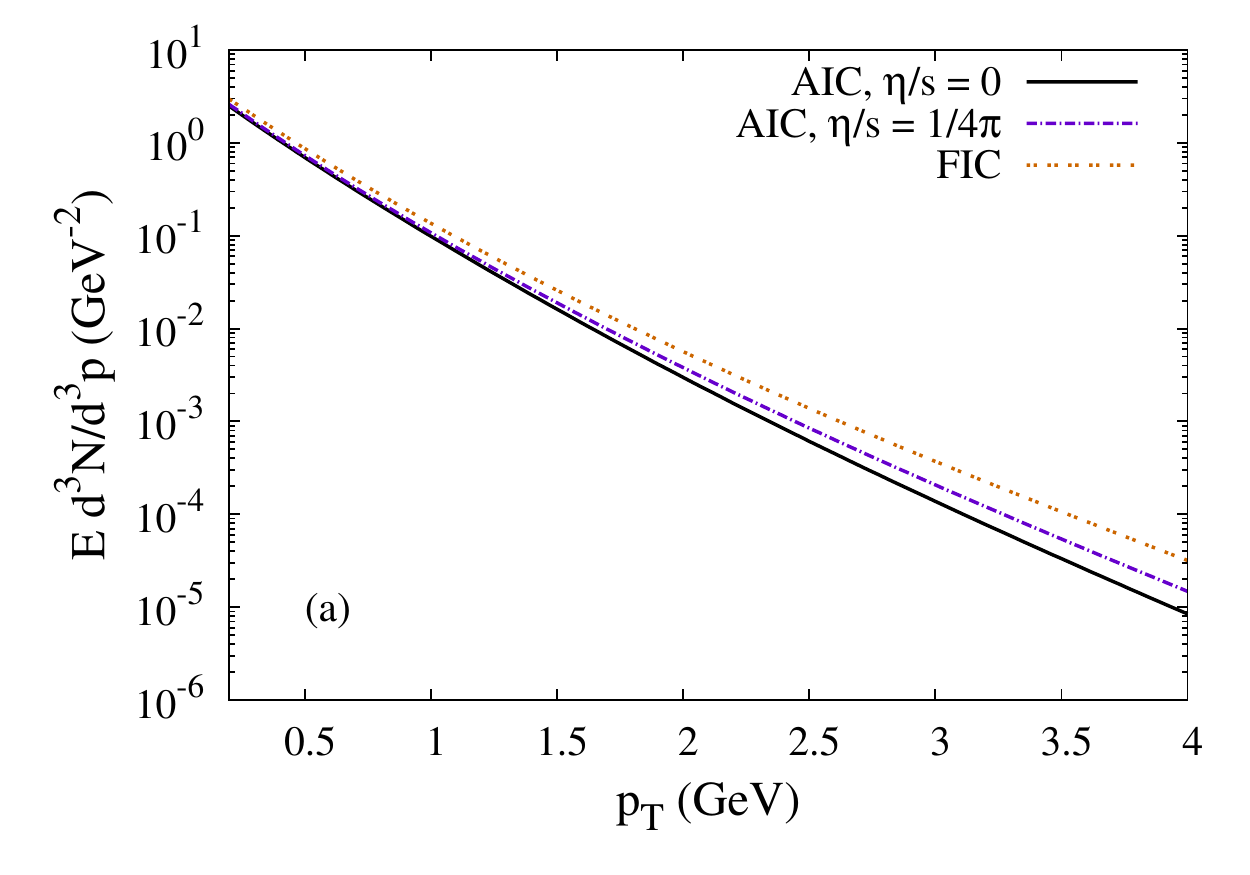}
\includegraphics[width=8.5cm]{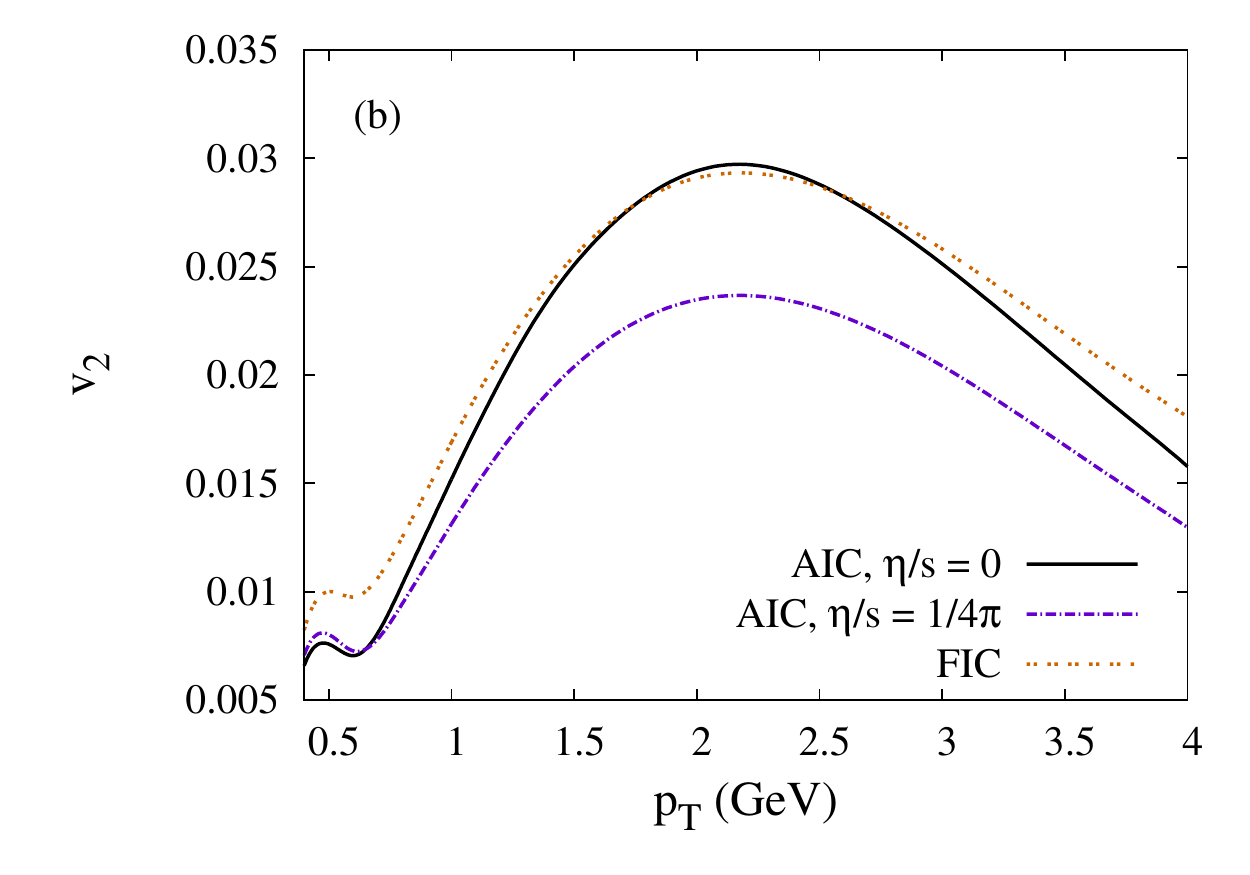}
\caption{(Color online)  The net thermal photon yield (left panel) and $v_2$
(right panel), showing the effect of FICs. Note that the curve labeled ``FIC'' also includes all
viscous corrections (time evolution and $\delta f$).}
\label{yield_v2_FIC}
\end{figure}
As done previously, only this time with FICs, we plot the thermal photon
$v_2$ for QGP and HG. This is shown in Figure \ref{v2_FIC}. Finally, the
net photon spectrum and $v_2$ are shown in Figure \ref{yield_v2_FIC}. Clearly, in the centrality range studies in this work, the hot spots and large gradients generated by the fluctuating initial conditions lead to a harder photon spectrum and to a larger elliptic flow, and this remains true with the inclusion of a finite shear viscosity to entropy density ratio.
\section{Conclusion}
\label{conclusion}

In this work we have sought to establish the quantitative importance of a finite shear viscosity coefficient and of fluctuating initial conditions on two real photon observables: the one-body spectrum and the transverse momentum dependence of the elliptic flow coefficient. This was done using \textsc{music},  a realistic 3+1D relativistic hydrodynamical simulation. Importantly, comparisons between cases with and without viscous corrections were done using conditions tuned to hadronic experimental data, and this was the case also for studies involving FICs. Results obtained here show that the combined effects of the viscosity and of the FICs are large enough to make their inclusion mandatory in any attempt to quantitatively extract transport coefficients of the hot and dense matter from thermal photon data. It was not the point of this work to explicitly compare with experimental measurements just yet. Firstly, 3+1D relativistic viscous hydrodynamics models are in their infancy, and systematic studies of {\it all} parameter dependences, in the spirit of that in Ref. \cite{Shen:2010uy} for example,  will be useful to establish a more precise quantitative  link between  observables  and the underlying hydrodynamics. Secondly, in what concerns the photon sources, an inclusive and consistent treatment of all of them (pQCD photons, photons from jets interacting and fragmenting while losing energy \ldots) with and without viscosity is still to be done. Finally, exploring  the consequences of what has been found here on electromagnetic observables at the LHC  should prove interesting and relevant. 

In closing, it is worth mentioning that recently the PHENIX collaboration at RHIC has extracted a direct photon $v_2$ from measured data \cite{Adare:2011zr}. Interestingly, this analysis concludes that the direct photon elliptic flow is comparable in magnitude to that of the $\pi^0$. This large photon elliptic flow is a challenge to most approaches, but may contain some clues about early dynamics prior to hadronic freeze-out \cite{vanHees:2011vb}.  However, a complete theoretical analysis including all that is known in the literature about the production of direct photons in relativistic heavy ion collisions (viscous effects, FICs, the effect of realistic 3+1D hydrodynamical modelling, hadronic chemical potentials, primordial flow) is needed before a more precise assessment can be obtained. 

\begin{acknowledgments}
This work was funded in part by the Natural Sciences and Engineering
Research Council of Canada, in part by the Fonds de recherche du Qu\'ebec -
Nature et technologies, in part by
 the US Department of Energy under DOE Contract No.DE-AC02-98CH10886, and
in part
by a Laboratory Directed Research and Development Grant from Brookhaven
Science Associates. J.-F. P. acknowledges a Hydro-Quebec Graduate Fellowship. 
\end{acknowledgments}

\appendix*
\section{Viscous photon rates}

Our starting point for the photon emission rate of a $(1+2 \rightarrow
3+\gamma)$ process is the  expression leading to Eq. (\ref{phot_rate}):
\begin{eqnarray}
E\frac{d^3R}{d^3p} &=
&\int\frac{d^3p_{1}}{2E_{1}(2\pi)^3}\frac{d^3p_{2}}{2E_{2}(2\pi)^3}\frac{d^3p_{3}}{2E_{3}(2\pi)^3}\frac{2\pi}{2}
|\mathcal M|^2\nonumber \\
&\times
&\delta^4(p_{1}+p_{2}-p_{3}-p)f(p_{1})f(p_{2})\left(1+f(p_{3})\right)\nonumber \\
\end{eqnarray}
The subscripts 1,2 and 3 refer to the two incoming particles and the
outgoing particle (not the photon) respectively.
Notice that the momentum distribution functions depend on the four-momentum
and not only on the zero component (the energy).
This can be written as (with $s=(p_1+p_2)^2,
t=(p_2-p)^2, u=-s-t+m_1^2+m_2^2+m_3^2$)
\begin{align}
E\frac{d^3R}{d^3p} &=
\frac{1}{2^{5}(2\pi)^{8}E^2}\int_{s^{\rm min}}^{\infty}ds\int_{t^{\rm min}}^{t^{\rm max}}dt\int_{E_{1}^{\rm min}}^{\infty}dE_{1}\int_{E_{2}^{\rm min}}^{E_{2}^{\rm max}}dE_{2}
\int_{0}^{2\pi}d\phi_{1}\int_{0}^{2\pi}d\phi_{2}|\mathcal M|^2 \nonumber \\
&\delta(s-m_{1}^{2}-m_{2}^{2}-2E_{1}E_{2}+2|p_{1}||p_{2}|\cos{\theta})f(p_{1})f(p_{2})(1+f(p_{1}+p_{2}-p))
\end{align}
where
\begin{align}
\cos{\theta} &= \cos{\theta_1}\cos{\theta_2} + \sin{\theta_1}\sin{\theta_2}
\cos{(\phi_2 - \phi_1)} \nonumber \\
\cos{\theta_1} &= \frac{-t-s+m_2^2+m_3^2+2EE_1}{2E|\vec{p}_1|}, \quad
\cos{\theta_2} = \frac{-t-m_2^2+2EE_2}{2E|\vec{p}_2|}
\end{align}
The $\theta$ angles represent the angle between a particle's three momentum
and the photon's. In other words,
\begin{equation}
\vec{p}_a\cdot\vec{p}_\gamma = |\vec{p}_a||\vec{p}_\gamma| \cos{\theta_a}
\end{equation}
The $\phi$ angles represent the azimuthal angles of the incoming particle's
direction around the photon's direction.
The integration boundaries are given by
\begin{eqnarray}
s^{\rm min} \geq (m_1+m_2)^2, \quad s^{\rm min} \geq m_3^2 \\ \nonumber \\
t^{\rm min(max)} = m_1^2 + m_3^2 -\frac{1}{2}\Big(
\frac{s+m_1^2-m_2^2}{\sqrt{s}} \Big)\Big( \frac{s+m_3^2}{\sqrt{s}} \Big)
\nonumber \\ -(+) \sqrt{\frac{(s+m_1^2-m_2^2)^2-4sm_1^2}{s}}\Big(
\frac{s-m_3^2}{2\sqrt{s}} \Big) \\ \nonumber \\
E_1^{\rm min} = \frac{Em_1^2}{m_1^2-u} + \frac{m_1^2-u}{4E} \\ \nonumber \\
\frac{-b + \sqrt{b^2 - ac}}{a} \leq E_2 \leq \frac{-b - \sqrt{b^2 - ac}}{a}
\quad \text{($a$ is negative)}
\end{eqnarray}

The $a$, $b$ and $c$ coefficients are given below. Now, we can take care of the $\phi_2$
integration with the Dirac delta. Its roots are
\begin{equation}
\phi_{\pm} = \phi_1 \pm \cos^{-1}{\Big(\frac{-s+m_1^2+m_2^2+2E_1E_2 -
2|\vec{p}_1||\vec{p}_2|\cos{\theta_1}\cos{\theta_2}}{2|\vec{p}_1||\vec{p}_2|
\sin{\theta_1}\sin{\theta_2}}\Big)} \label{dirac_roots}
\end{equation}

However, now we need to keep in mind that where we had $f(E_2)$ for the
ideal (no viscous corrections)
case, we now have $f(E_2,\theta_2,\phi_2)$, so the $\phi_2$ integration
goes like
\begin{eqnarray}
\int_{0}^{2\pi}d\phi_{2}\delta(s-m_{1}^{2}-m_{2}^{2}-2E_{1}E_{2}+2|p_{1}||p_{2}|\cos{\theta})f(E_{2},\theta_{2},\phi_{2})= \nonumber \\
\int_{0}^{2\pi}d\phi_{2}\sum_{j=\pm}\frac{\delta(\phi_{2}-\phi_{j})f(E_{2},\theta_{2},\phi_{2})}{2|\vec{p}_{1}||\vec{p}_{2}|\sin{\theta_{1}}\sin{\theta_{2}}
\sin{(\phi_{j}-\phi_{1})}} = \nonumber \\
\frac{1}{2|\vec{p}_{1}||\vec{p}_{2}|\sin{\theta_{1}}\sin{\theta_{2}}}\Big(\frac{f(E_{2},\theta_{2},\phi_{+})}{\sqrt{1-\cos^{2}{(\phi_{+}-\phi_{1})}}}+
\frac{f(E_{2},\theta_{2},\phi_{-})}{\sqrt{1-\cos^{2}{(\phi_{-}-\phi_{1})}}}\Big) =\nonumber \\
\frac{1}{2|\vec{p}_{1}||\vec{p}_{2}|\sin{\theta_{1}}\sin{\theta_{2}}}\frac{1}{\sqrt{1-\cos^{2}{(\phi_{+}-\phi_{1})}}}\Big(f(E_{2},\theta_{2},\phi_{+})
+f(E_{2},\theta_{2},\phi_{-})\Big)=\nonumber\\
\frac{E}{\sqrt{aE_{2}^2+2bE_{2}+c}}\Big(f(E_{2},\theta_{2},\phi_{+})+f(E_{2},\theta_{2},\phi_{-})\Big) \label{phi2_int_visc}
\end{eqnarray}
where the $a,b$ and $c$ coefficients are given by
\begin{eqnarray}
a &=& -(s+t-m_2^2 -m_2^3)^2 \nonumber \\
b &=& E \big( (s+t-m_2^2-m_2^3)(s-m_1^2-m-2^2)-2m_1^2(m_2^2-t) \big)
\nonumber \\ &&+E_1(m_2^2 - t)(s+t-m_2^2-m_2^3) \nonumber \\
c &=& c_2E_1^2 + c_1E_1 + c_0 \nonumber \\
c_2 &=& -(t-m_2^2)^2 \nonumber \\
c_1 &=& -2E\big( 2m_2^2(s+t-m_1^2-m_2^2)-(m_2^2-t)(s-m_1^2-m_2^2) \big)
\nonumber \\
c_0 &= & 4E^2m_1^2m_2^2 + m_2^2(s+t-m_2^2-m_2^3) + m_1^2(m_2^2-t)^2 -
E^2(s-m_1^2-m_2^2)^2 \nonumber \\&&+
(s-m_1^2-m_2^2)(t-m_2^2)(s+t-m_2^2-m_2^3) \label{abc_coeff}
\end{eqnarray}

So at this point, the rate is given by
\begin{align}
E\frac{d^3R}{d^3p}&=\frac{1}{32(2\pi)^8E}\int_{s^{\rm min}}^{\infty}ds\int_{t^{\rm min}}^{t^{\rm max}}dt\int_{E_{1}^{\rm min}}^{\infty}dE_{1}\int_{E_{2}^{\rm min}}^{E_{2}^{\rm max}}dE_{2}
\int_{0}^{2\pi}d\phi_{1}|\mathcal M|^{2}f(E_{1},\theta_{1},\phi_{1})
\nonumber \\
&\Big(f(E_{2},\theta_{2},\phi_{+})+f(E_{2},\theta_{2},\phi_{-})\Big)(1+f(E_{1}+E_{2}-E))\frac{1}{\sqrt{aE_{2}^2+2bE_{2}+c}}
\end{align}
We have not corrected the distribution functions for the outgoing particle
(Pauli blocking or Bose enhancement). At
this point, we look more carefully at the form of the correction (choosing Bose enhancement for illustrative purposes):
\begin{eqnarray}
f(p_{a})=f_{0}(E_{a})+\delta f(p_{a})= 
f_{0}(E_{a})+\Big(\frac{\eta}{s}\frac{1}{2T^3}f_{0}(E_{a})\big(1+f_{0}(E_{a})\big)p_{a}^{\alpha}p_{a}^{\beta}\frac{\pi_{\alpha\beta}}{\eta}\Big)
\end{eqnarray}
One needs to properly express $p^\alpha$ in terms of  integration
variables. Because the incoming particles angles are given with respect to
the
photon's direction, the angles of the photon (in the fluid frame) will
explicitly appear in the expressions for the particles four momentum. The
zero component
of $p^\alpha$ is just the energy, so the interesting parts are the 1,2 and
3 components. In what follows, the subscript $a$ will refer to the incoming
particles (1 or 2), and the subscript $\gamma$ will denote angles for the
photon. The photon angles are given with respect to the local frame of
each
fluid cell. Note that even if we use the $\phi_a$ notation, $\phi_2$ is
still given by equation \eqref{dirac_roots}. Also, all 3-vectors in the
following expressions are unitary vectors. The photon's direction is given
by
\begin{equation}
\vec{p_{\gamma}}=(\sin{\theta_{\gamma}}\cos{\phi_{\gamma}},\sin{\theta_{\gamma}}\sin{\phi_{\gamma}},\cos{\theta_{\gamma}})
\end{equation}
We need to specify the origin for $\phi_a$ ($\phi_a=0$). Since the $\phi_a$
integration is over $2\pi$, the origin is
arbitrary. Let's choose the origin to be in the z-$p_\gamma$ plane. Then,
\begin{align}
\vec{p_{a}}(\phi_{a}=0)&=\big(\sin{(\theta_{\gamma}-\theta_{a})}\cos{\phi_{\gamma}},\sin{(\theta_{\gamma}-\theta_{a})}\sin{\phi_{\gamma}},\cos{(\theta_{\gamma}-\theta_{a})}\big) \nonumber \\
&=\big((\sin{\theta_{\gamma}}\cos{\theta_{a}}-\cos{\theta_{\gamma}}\sin{\theta_{a}})\cos{\phi_{\gamma}}, \nonumber \\
&(\sin{\theta_{\gamma}}\cos{\theta_{a}}-\cos{\theta_{\gamma}}\sin{\theta_{a}})\sin{\phi_{\gamma}},
\cos{\theta_{\gamma}}\cos{\theta_{a}}+\sin{\theta_{\gamma}}\sin{\theta_{a}}\big)
\end{align}
To have an expression for $\vec{p_{a}}$, we need to rotate
$\vec{p_{a}}(\phi_{a}=0)$ by an angle of $\phi_{a}$ around
$\vec{p_{\gamma}}$. The rotation matrix
is then given by
\begin{equation}
R=\left(
\begin{array}{ccc}
r_{11} & r_{12} & r_{13} \\
r_{21} & r_{22} & r_{23}\\
r_{31} & r_{32} & r_{33}
\end{array}
\right)
\end{equation}
\begin{align}
r_{11} & =
\cos{\phi_{a}}+\sin^{2}{\theta_{\gamma}}\cos^{2}{\phi_{\gamma}}(1-\cos{\phi_{a}}) \nonumber \\
r_{12} & =
\sin^{2}{\theta_{\gamma}}\cos{\phi_{\gamma}}\sin{\phi_{\gamma}}(1-\cos{\phi_{a}})-\cos{\theta_{\gamma}}\sin{\phi_{a}} \nonumber \\
r_{13} & =
\sin{\theta_{\gamma}}\cos{\theta_{\gamma}}\cos{\phi_{\gamma}}(1-\cos{\phi_{a}})+\sin{\theta_{\gamma}}\sin{\phi_{\gamma}}\sin{\phi_{a}} \nonumber \\
r_{21} & =
\sin^{2}{\theta_{\gamma}}\cos{\phi_{\gamma}}\sin{\phi_{\gamma}}(1-\cos{\phi_{a}})+\cos{\theta_{\gamma}}\sin{\phi_{a}} \nonumber \\
r_{22} & =
\cos{\phi_{a}}+\sin^{2}{\theta_{\gamma}}\sin^{2}{\phi_{\gamma}}(1-\cos{\phi_{a}}) \nonumber \\
r_{23} & =
\sin{\theta_{\gamma}}\cos{\theta_{\gamma}}\sin{\phi_{\gamma}}(1-\cos{\phi_{a}})-\sin{\theta_{\gamma}}\cos{\phi_{\gamma}}\sin{\phi_{a}} \nonumber \\
r_{31} & =
\sin{\theta_{\gamma}}\cos{\theta_{\gamma}}\cos{\phi_{\gamma}}(1-\cos{\phi_{a}})-\sin{\theta_{\gamma}}\sin{\phi_{\gamma}}\sin{\phi_{a}} \nonumber \\
r_{32} & =
\sin{\theta_{\gamma}}\cos{\theta_{\gamma}}\sin{\phi_{\gamma}}(1-\cos{\phi_{a}})+\sin{\theta_{\gamma}}\cos{\phi_{\gamma}}\sin{\phi_{a}} \nonumber \\
r_{33} & = \cos{\phi_{a}}+\cos^{2}{\theta_{\gamma}}(1-\cos{\phi_{a}})
\end{align}
So our expression for $\vec{p}_a$ is finally
\begin{equation}
\left[
\begin{array}{c}
p^{x}_{a} \\ p^{y}_{a} \\ p^{z}_{a}
\end{array}
\right]=R
\left[
\begin{array}{c}
(\sin{\theta_{\gamma}}\cos{\theta_{a}}-\cos{\theta_{\gamma}}\sin{\theta_{a}})\cos{\phi_{\gamma}} \\ (\sin{\theta_{\gamma}}\cos{\theta_{a}}-\cos{\theta_{\gamma}}\sin{\theta_{a}})\sin{\phi_{\gamma}} \\ \cos{\theta_{\gamma}}\cos{\theta_{a}}+\sin{\theta_{\gamma}}\sin{\theta_{a}}
\end{array}
\right]
\end{equation}
Note that all the components of $p_a$ are proportional to either one or no power of
$\cos{\theta_{a}}$, $\sin{\theta_{a}}$, $\cos{\phi_{a}}$ and
$\sin{\phi_{a}}$, so we can rewrite, for instance, the $x$ component as (with the A,B
and C coefficients that only depend on  photon angles)
\begin{eqnarray}
p^x_{a} =
\cos{\theta_{a}}\Big\{\cos{\phi_{a}}A^c_x+\sin{\phi_{a}}B^c_x+C^c_x\Big\} +
\sin{\theta_{a}}\Big\{\cos{\phi_{a}}A^s_x+\sin{\phi_{a}}B^s_x+C^s_x\Big\}
\end{eqnarray}
and in a similar way we have
\begin{eqnarray}
p^y_{a} =
\cos{\theta_{a}}\Big\{\cos{\phi_{a}}A^c_y+\sin{\phi_{a}}B^c_y+C^c_y\Big\} +
\sin{\theta_{a}}\Big\{\cos{\phi_{a}}A^s_y+\sin{\phi_{a}}B^s_y+C^s_y\Big\}\\
p^z_{a} =
\cos{\theta_{a}}\Big\{\cos{\phi_{a}}A^c_z+\sin{\phi_{a}}B^c_z+C^c_z\Big\} +
\sin{\theta_{a}}\Big\{\cos{\phi_{a}}A^s_z+\sin{\phi_{a}}B^s_z+C^s_z\Big\}
\end{eqnarray}
Now, going back to the expression for the rate, only the viscous corrected
momentum distribution functions actually depend on $\phi_{1}$ so
let's look at the $\phi_{1}$ integration of the viscous corrections
\begin{eqnarray}
\int^{2\pi}_{0}d\phi_{1}\Big(f_{0}(E_{1})+\delta
f(p_{1})\Big)\Big(f_{0}(E_{2})+\delta f(p_{2})\Big) \nonumber \\
\approx\int^{2\pi}_{0}d\phi_{1}f_{0}(E_{1})f_{0}(E_{2})+f_{0}(E_{1})\delta
f(p_{2})+f_{0}(E_{2})\delta f(p_{1})
\end{eqnarray}
where  the term proportional to $\delta f(p_1) \delta
f(p_2)$ has been neglected. Since $f_0(E_a)$ is independent of $\phi_a$, what we really need
to look at is
\begin{equation}
\int^{2\pi}_{0}d\phi_{1}\delta
f(p_{a})\propto\int^{2\pi}_{0}d\phi_{1}p^{\alpha}_{a}p^{\beta}_{a}\frac{\pi_{\alpha\beta}}{\eta}
\end{equation}
It should be pointed out that $\phi_{2}$ (which is given by
$\phi_{+}$ or $\phi_{-}$) differs from $\phi_{1}$ by a term which, as far as the integration over $\phi_1$ is concerned, is a constant.  Because the only
dependance on $\phi_{2}$ comes from sine or cosine terms, and because the
$\phi_1$ integration is over a full cycle (from 0 to $2\pi$), $\phi_{2}$ is
actually the same as $\phi_{1}$. Or, more precisely,
\begin{multline}
\int^{2\pi}_{0}d\phi_{1}f(\cos{\phi_{1}},\sin{\phi_{1}})=\int^{2\pi}_{0}d\phi_{1}f(\cos{\phi_{2}},\sin{\phi_{2}}) \\
=\int^{2\pi}_{0}d\phi_{1}f(\cos{\phi_{+}},\sin{\phi_{+}})=\int^{2\pi}_{0}d\phi_{1}f(\cos{\phi_{-}},\sin{\phi_{-}})
\end{multline}
One can replace the
$\Big(f(E_{2},\theta_{2},\phi_{+})+f(E_{2},\theta_{2},\phi_{-})\Big)$ term
in \eqref{phi2_int_visc} by $2f(E_{2},
\theta_{2},\phi_{1})$. Furthermore, the $\phi_1$ integration will then be
treated the same way regardless of the particle being considered (1 or
2).
Considering, for instance, the $p^{x}_{1}p^{y}_{1}$ term and carrying 
out the $\phi_{1}$ integration, one obtains 
\begin{multline}
\int^{2\pi}_{0}d\phi_{1}p^{x}_{1}p^{y}_{1} = \int^{2\pi}_{0}d\phi_{1}\Big(\cos{\theta_{1}}\big\{\cos{\phi_{1}}A^c_x+\sin{\phi_{a}}B^c_x+C^c_x\big\} + \sin{\theta_{1}}\big\{\cos{\phi_{1}}A^s_x+\sin{\phi_{1}}B^s_x+C^s_x\big\}\Big)\\ \Big(\cos{\theta_{1}}\big\{\cos{\phi_{1}}A^c_y+\sin{\phi_{1}}B^c_y+C^c_y\big\} + \sin{\theta_{1}}\big\{\cos{\phi_{1}}A^s_y+\sin{\phi_{1}}B^s_y+C^s_y\big\}\Big)\\
=\pi\Big[\cos^2{\theta_{1}}\Big(A^c_{x}A^c_{y}+B^c_{x}B^c_{y}+2C^c_{x}C^c_{y}\Big)+\sin^2{\theta_{1}}\Big(A^s_{x}A^s_{y}+B^s_{x}B^s_{y}+2C^s_{x}C^s_{y}\Big)\\+\cos{\theta_{1}}\sin{\theta_{1}}\Big(A^c_{x}A^s_{y}+B^c_{x}B^s_{y}+2C^c_{x}C^s_{y}+A^s_{x}A^c_{y}+B^s_{x}B^c_{y}+2C^s_{x}C^c_{y}\Big)\Big] \label{palpha_pbeta_expansion}
\end{multline}
If we were dealing with particle 2, the $\theta_{1}$'s would be
$\theta_{2}$'s, and if we were looking at other x-y-z combinations the A,B
and C coefficients would be different, but the general form is always the
same. The important thing is that, after the $\phi_1$ integration, the
$p_a^\alpha p_a^\beta$ terms can easily
be split into a part that depends only on E and T, the photon energy and the local temperature (via the $\cos{\theta_a}$
and $\sin{\theta_a}$), and one that depends only on the photon angles (the
A,B and C
coefficients). So in the end, the photon production rate, including
momentum distribution function modifications from viscosity, is given by
\begin{multline}
E\frac{d^3R}{d^3p}=\frac{1}{16(2\pi)^8E}\int_{s^{\rm min}}^{\infty}ds\int_{t^{\rm min}}^{t^{\rm max}}dt\int_{E_{1}^{\rm min}}^{\infty}dE_{1}\int_{E_{2}^{\rm min}}^{E_{2}^{\rm max}}dE_{2}
|\mathcal M|^{2}(1+f_{0}(E_{1}+E_{2}-E))\\
\times \frac{1}{\sqrt{aE_{2}^2+2bE_{2}+c}}
\Big(f_{0}(E_{1})f_{0}(E_{2})+\frac{\eta}{s}\frac{1}{2T^3}\big[f_{0}(E_{2})f_{0}(E_{1})\big(1+f_{0}(E_{1})\big)p^{\alpha}_{1}p^{\beta}_{1}\\ + f_{0}(E_{1})f_{0}(E_{2})\big(1+f_{0}(E_{2})\big)p^{\alpha}_{2}p^{\beta}_{2}\big]
\frac{\pi_{\alpha\beta}}{\eta}\Big)
\label{visc_rate}
\end{multline}
where it is understood that the $p^{\alpha}_{a}p^{\beta}_{a}$ terms have
been integrated over $\phi_{1}$, from 0 to $2\pi$. 
{ Now if we substitute Eq. (\ref{palpha_pbeta_expansion}) (and similar expressions for other $p_a^\alpha p_a^\beta$ combinations) into
the above expression, the result can be expressed in terms of four-integrals for which the result depends only on E and T, multiplied
by a combination of terms that depend only on $\theta_\gamma$ and $\phi_\gamma$. }
The integrals that need to be computed are then of the type 
\begin{eqnarray}
\int_{s^{\rm min}}^{\infty}ds\int_{t^{\rm min}}^{t^{\rm max}}dt\int_{E_{1}^{\rm min}}^{\infty}dE_{1}\int_{E_{2}^{\rm min}}^{E_{2}^{\rm max}}dE_{2}|\mathcal M|^{2} 
(1+f_{0}(E_{1}+E_{2}-E))\frac{1}{\sqrt{aE_{2}^2+2bE_{2}+c}}\big[...\big]
\end{eqnarray}
where $\big[...\big]$ is one of
\begin{eqnarray}
\big[...\big]=\left\{
\begin{array}{l}
f_{0}(E_{1})(E_2^2-m_2^2)\cos^2{\theta_{2}}f_{0}(E_{2})\big(1+f_{0}(E_{2})\big) \\ \\
f_{0}(E_{1})(E_2^2-m_2^2)\sin^2{\theta_{2}}f_{0}(E_{2})\big(1+f_{0}(E_{2})\big) \\ \\
f_{0}(E_{1})(E_2^2-m_2^2)\cos{\theta_{2}}\sin{\theta_{2}}f_{0}(E_{2})\big(1+f_{0}(E_{2})\big) \\ \\
f_{0}(E_{2})(E_1^2-m_1^2)\cos^2{\theta_{1}}f_{0}(E_{1})\big(1+f_{0}(E_{1})\big) \\ \\
f_{0}(E_{2})(E_1^2-m_1^2)\sin^2{\theta_{1}}f_{0}(E_{1})\big(1+f_{0}(E_{1})\big) \\ \\
f_{0}(E_{2})(E_1^2-m_1^2)\cos{\theta_{1}}\sin{\theta_{1}}f_{0}(E_{1})\big(1+f_{0}(E_{1})\big)
\end{array} \right.
\label{sixterms}
\end{eqnarray}
where $(E_a^2-m_a^2)$ is just $|\vec{p}_a|^2$, which we have omitted in the
derivations above. { For instance, to compute the contribution to the correction from the $p^{x}_{1}p^{y}_{1}$ term, one would
need the last three terms in Eq. (\ref{sixterms})  (the ones corresponding to $\cos^2{\theta_{1}}$, $\sin^2{\theta_{1}}$, 
$\cos{\theta_{1}}\sin{\theta_{1}}$). Each of these terms, would then be multiplied by some combination of the $A$, $B$, and $C$ terms.
In this example, the $\cos^2{\theta_{1}}$ gets multiplied by $\Big(A^c_{x}A^c_{y}+B^c_{x}B^c_{y}+2C^c_{x}C^c_{y}\Big)$, as in Eq.  (\ref{palpha_pbeta_expansion}.)}

\bibliography{hydro}

\end{document}